\documentclass[aps,pra,showpacs,superscriptaddress]{revtex4}

\usepackage{graphicx}
\usepackage{bm}
\usepackage{amssymb}

\def\norc #1 {\ {\Vert} #1 {\Vert_{c}}}
\def\norN #1 {\ {\Vert} #1 {\Vert_{N}}}
\def\nor #1 {\ {\Vert} #1 {\Vert}}

\def \Ninf#1 {  {\Vert {#1} \Vert_\infty } }
\def \NLip#1 {  {\Vert {#1} \Vert_\theta } }
\def \Norm#1 {  {\Vert {#1} \Vert } }

\def \bra{{ \langle}}
\def \ket{{ \rangle}}

\def\cL{{\mathcal{L}}}
\def\cK{{\mathcal{K}}}

\def\cG{{\mathcal{G}}}
\def\cF{{\mathcal{F}}}
\renewcommand{\Re}{\mathrm{\,Re}}

\newcommand{\beq}{\begin{equation}}
\newcommand{\eeq}{\end{equation}}
\newcommand{\barr}{\begin{eqnarray}}
\newcommand{\earr}{\end{eqnarray}}

\newcommand{\andy}[1]{ }

\begin{document}

\def\figwidth{8.5cm}


\title{Control of decoherence: analysis and comparison of
three different strategies}



\author{P. Facchi}
\affiliation{Dipartimento di Fisica, Universit\`a di Bari  I-70126
Bari, Italy \\ and Istituto Nazionale di Fisica Nucleare, Sezione
di Bari, I-70126 Bari, Italy}
\author{S. Tasaki}
\affiliation{Department of Applied Physics and Advanced  Institute
for Complex Systems, Waseda University, Tokyo 169--8555, Japan}
\author{S. Pascazio}
\affiliation{Dipartimento di Fisica, Universit\`a di Bari  I-70126
Bari, Italy \\ and Istituto Nazionale di Fisica Nucleare, Sezione
di Bari, I-70126 Bari, Italy}
\author{H. Nakazato}
\affiliation{Department of Physics, Waseda University, Tokyo
169--8555, Japan}
\author{A. Tokuse}
\affiliation{Department of Applied Physics and Advanced  Institute
for Complex Systems, Waseda University, Tokyo 169--8555, Japan}
\author{D. A. Lidar}
\affiliation{Chemical Physics Theory Group, Chemistry Department,
  University of Toronto, 80 St.\ George Street, Toronto, Ontario M5S 3H6, Canada}



\date{\today}

\begin{abstract}
We analyze and compare three different strategies, all aimed at
controlling and eventually halting decoherence. The first strategy
hinges upon the quantum Zeno effect, the second makes use of
frequent unitary interruptions (``bang-bang" pulses and their
generalization, quantum dynamical decoupling), and the third of a
strong, continuous coupling. Decoherence is shown to be suppressed
only if the frequency $N$ of the measurements/pulses is large
enough or if the coupling $K$ is sufficiently strong. Otherwise,
if $N$ or $K$ are large, but not extremely large, all these
control procedures accelerate decoherence. We investigate the
problem in a general setting and then consider some practical
examples, relevant for quantum computation.
\end{abstract}



\maketitle


\section{Introduction}

Interactions with the environment deteriorate the purity of
quantum states. This general phenomenon, known as decoherence
\cite{decoherencereview}, is a serious obstacle against the
preservation of quantum superpositions and entanglement over long
periods of time. Decoherence entails non-unitary evolutions, with
serious consequences, like a loss of information and/or
probability leakage towards the environment.

This issue is recently attracting much attention in view of
interesting applications: for instance, the possibility of
controlling and eventually halting decoherence is a key problem in
quantum computation \cite{Review}, where several computational
states are simultaneously described by a single wave function and
parallel information processing is carried out by unitary
operations. In such a situation, efficient quantum algorithms need
large scale computations, performed over (microscopically) long
time spans \cite{Unruh}.

A number of interesting schemes have been proposed during the last
few years in order to counter the effects of decoherence. Among
these, there are quantum error correcting codes
\cite{ErrorCorrecting}, schemes based on feedback
or stochastic control \cite{feedback}, the use of decoherence free
subspaces and noiseless subsystems
\cite{NoiselessSub} and mechanisms based on ``bang-bang" (BB)
pulses and their generalization, quantum dynamical decoupling
\cite{BBDDsem,BBDD,Zanardi,Vitali,BL}. In this context, it was
recently proposed \cite{bang} that the method of dynamical
decoupling can be unified with the basic ideas underlying the
quantum Zeno effect (QZE)
\cite{QZEseminal,MS} (for a review, see
\cite{QZEreview,PIO}). In particular, the decoherence free subspace
is one of the dynamically generated quantum Zeno subspaces
\cite{FPsuper}, within which the dynamics is not trivial
\cite{compactregularize} and whose subtle mathematical aspects are
still debated
\cite{Friedman72,MS,Gustavson,ExnerIchinose,Schmidt,MisraAntoniou}.

It is worth stressing that the ``bang-bang" scheme is a
well-established ``classical" control method, typically used in
engineering problems and in connection with spin-echo techniques:
see, for instance, Ref. \cite{classicalBB}. Its revival in
quantum-information-related problems is only very recent. The key
ingredient of BB and dynamical decoupling is to apply frequent
(unitary) interruptions during the evolution of the system, in
order to suppress the system-environment interaction. The
similarity with the QZE was already noticed in \cite{BBDDsem}. It
is however clear that the two procedures are physically
equivalent, if one adheres to the commonly accepted interpretation
of the QZE as a \emph{bona fide} dynamical process, that can be
completely explained in terms of \emph{unitary} evolutions
\cite{QZEbxl}. One should notice that this idea hinges upon a
seminal remark by Wigner \cite{Wigner63}, who introduced in 1963
the notion of ``spectral decomposition," namely a dynamical
process that associates a different wave packet with each
eigenvalue of the observable to be measured. For example, the
interesting proposal by Cook \cite{Cook} and the subsequent
experiment with Rabi oscillations \cite{WinelandZeno} can be
easily interpreted in fully dynamical terms when one observes that
the ``measurement" was realized as a dynamical process (optical
pulse irradiation) \cite{PIO,QZEbxl,Itanodisc}.

Once this physical equivalence is appreciated, the next logical
step is a natural one: after having analyzed and understood the
consequences of frequent unitary pulses, one studies the effect of
a strong (unitary) continuous coupling. The relationship between
these two procedures can be made mathematically precise (see Sec.\
\ref{sec.general}) and is of interest in itself: if an external
field or ``apparatus" is coupled to the system in such a way that
the state of the system is ``monitored" in some sense
\cite{Simonius,HS,Peres,Schulman98}, a Zeno-like dynamics takes
place in the strong coupling limit and once again one can tailor
decoherence-free subspaces \cite{bang}. This happens to be one of
the most efficient and convenient control procedures, from a
practical point of view.

The aim of this article is to investigate these different physical
procedures (Zeno, BB-dynamical decoupling and continuous coupling)
and compare their effects. We will study the dynamics generated by
very frequent interruptions (projective measurements or unitary
``kicks," yielding dynamical decoupling), or by very strong
coupling, and investigate the possibility of designing
decoherence-free subspaces. The method is general and can be
applied to diverse situations of practical interest, such as atoms
and ions in cavities, organic molecules, quantum dots and
Josephson junctions
\cite{Wineland,expt,qdot,JJs}.

Our main objective is to endeavor to understand whether it is
possible to \emph{control} decoherence
\cite{Calarco,Falci,Paladino:02,Falci:03,KofKu}. Clearly, this
requires a thorough understanding of the physical mechanisms that
provoke decoherence and in general dissipative phenomena. One
finds that very frequent kicks/measurements or very strong
couplings can indeed control the evolution of the system and
suppress decoherence. The physical mechanisms at the origin of
this phenomenon are very close to the quantum Zeno effect.
However, if the kicks/measurements are not extremely frequent or
the coupling not extremely strong, both controls may {\it
accelerate} decoherence. This extends the notion of ``inverse"
quantum Zeno effect (IZE) \cite{IZE,Heraclitus} to a wider
framework (not necessarily based on projection operators and
non-unitary dynamics) and entails a deterioration of the
performance of these schemes. We will analyze this effect in great
detail and see that in order to avoid it, one must carefully
design the control and study the timescales involved. Our analysis
is of general validity; however, for the sake of definiteness, we
will study in particular the control of thermal decoherence
\cite{TTFP}.

This article is organized as follows. In Sec.\
\ref{sec.general} we briefly review the main features of the
different control procedures. Our analysis is based on a master
equation which is derived in Sec.\ \ref{sec.model}, where the
relevant timescales are emphasized and the general type of
interaction specified. We then consider the case of thermal
decoherence, discussing the Zeno control, the control via
dynamical decoupling and the control by means of a strong
continuous coupling in Secs.\
\ref{sec.Zeno}, \ref{sec.kickb} and \ref{sec:dyndec}, respectively.
Some relevant examples are then considered in Sec.\
\ref{sec:examples}, where we focus on the primary role of
the form factors of the interaction in order to compare the
different control procedures. Section
\ref{sec.conclusions} is devoted to conclusions and perspectives.

\section{Control procedures: generalities}
\label{sec.general}

Let the total system consist of a target system and a reservoir
and its Hilbert space ${\cal H}_{\rm tot}={\cal H}_S \otimes {\cal
H}_B$ be expressed as the tensor product of the system Hilbert
space ${\cal H}_S$ and the reservoir Hilbert space ${\cal H}_B$.
The total Hamiltonian
\begin{equation}
H_{\rm tot} =H_0 + H_{SB} = H_S \otimes \openone_B + \openone_S
\otimes H_B+ H_{SB} \label{Hamiltonian1}
\end{equation}
is the sum of the system Hamiltonian $H_S \otimes \openone_B$, the
reservoir Hamiltonian $\openone_S \otimes H_B$ and their
interaction $H_{SB}$, which is responsible for decoherence; the
operators $\openone_S$ and $\openone_B$ are the identity operators
in the Hilbert spaces ${\cal H}_S$ and ${\cal H}_B$, respectively,
and the operators $H_S$ and $H_B$ act on ${\cal H}_S$ and ${\cal
H}_B$, respectively.

The dynamics of the total system is conveniently expressed in
terms of the Liouville operator (Liouvillian) ${\cal L}_{\rm
tot}$, defined by
\begin{equation}\label{eq:Ltotdef}
{\cal L}_{\mathrm{tot}}\rho \equiv -i[H_{\rm tot},\rho]=
-i\left(H_{\rm tot} \rho - \rho H_{\rm tot}\right) \ ,
\end{equation}
where $\rho$ is the density matrix. If the Hamiltonian is given by
(\ref{Hamiltonian1}), the Liouvillian is accordingly decomposed
into
\begin{equation}
\label{Ltotdef}
\cL_{\mathrm{tot}}= \cL_0 + \cL_{SB}=\cL_S+\cL_B +\cL_{SB},
\end{equation}
where the meaning of the symbols is obvious. We will not
explicitly write the coupling constant $\lambda$ multiplying the
interaction Liouvillian $\cL_{SB}$.

We focus on a proper subspace ${\cal H}_{\rm comp} \subset {\cal
H}_S$, in which quantum computation is to be performed. For this
reason we will look in detail at the case
\begin{equation}\label{eq:directsum}
{\cal H}_S={\cal H}_{\rm comp}\oplus{\cal H}_{\rm orth} .
\end{equation}
In particular, when we will look at some concrete examples, in
Sec.\ \ref{sec:examples}, the computation subspace will be a
qubit, ${\cal H}_{\rm comp}=\mathbb{C}^2$.

Since, in general, the reservoir state is mixed, it is convenient
to describe the time evolution in terms of density matrices. In
the case of a quantum state manipulation, the initial state of the
total system $\rho(0)$ is set to be a tensor product of the system
initial state $\sigma(0)$ and a reservoir (usually equilibrium)
state $\rho_B$
\begin{equation}\label{eq:rhoprod}
\rho(0)=\sigma(0)\otimes \rho_B.
\end{equation}
The validity of this assumption, usually taken for granted, is
discussed in Appendix \ref{sec:A} (see also
\cite{factorassumption}). The system state $\sigma(t)$ at time $t$
is given by the partial trace of the state $\rho(t)$ of the whole
system with respect to the reservoir degrees of freedom:
\begin{equation}\label{eq:sigtrace}
\sigma(t) \equiv {\rm tr}_B \rho(t) .
\end{equation}
When $\sigma(t)$ is not unitarily equivalent to $\sigma(0)$ for a
given class of initial states, decoherence is said to occur. The
purpose of the control is to suppress such decoherence. Note that,
for the control of decoherence, it is not necessary to look at all
possible states: rather, it is sufficient to consider only those
initial states which are relevant to the quantum state
manipulation in question.

\subsection{Quantum Zeno control}

We first look at the Zeno control, by adapting the argument of
Ref.\ \cite{FPsuper}. The control is obtained by performing
frequent measurements of the system. The measurement is described
by a projection superoperator $\hat P$ acting on the density
matrix
\begin{equation}
\rho \to {\hat P}\rho \equiv \sum_n \left(P_n\otimes \openone_B
\right) \rho \left(P_n\otimes \openone_B\right),
\label{measurement}
\end{equation}
where $\{ P_n\}$ is a set of orthogonal projection operators
acting on ${\cal H}_S$. In the following, we restrict our analysis
to a measuring apparatus that does not ``select'' the different
outcomes (nonselective measurement) \cite{Schwinger}, with a
complete set of projection operators $\sum_n P_n=\openone_S$. The
measurement is designed so that
\begin{equation}\label{eq:C}
{\hat P}H_{SB} = \sum_n \left(P_n\otimes \openone _B\right) H_{SB}
\left(P_n\otimes \openone_B\right) =0.
\end{equation}
In terms of the Liouvillian, this condition reads
\begin{equation}\label{eq:C1}
{\hat P}\cL_{SB} {\hat P} = 0.
\end{equation}
(We will see in the next subsection that a similar requirement is
necessary for the BB control and for the control via a continuous
coupling.) The Zeno control consists in performing repeated
nonselective measurements at times $t=k \tau$ ($k=0,1,2,\ldots$)
(we include an initial ``state preparation" at $t=0$). Between
successive measurements, the system evolves via $H_{\rm tot}$. The
density matrix after $N+1$ measurements, with an initial state
$\rho(0)$, is given by
\begin{equation}
\rho(t)=\rho(N\tau) = \left({\hat P} e^{\cL_{\rm tot} \tau} {\hat
P}\right)^N   \rho(0)  \ .
\end{equation}
We take the limit $\tau\to 0$ while keeping $t=N \tau$ constant
and get
\begin{equation}
\rho(t) = {\hat P}\left[ 1 + {\hat P} {\cal L}_{\rm tot}{\hat P}
\tau + {\rm O}\left(\tau^2\right)\right]^{t\over\tau}  \rho(0)
\stackrel{\tau \to 0}{\longrightarrow} {\hat P} e^{{\hat P}{\cal
L}_{\rm tot}{\hat P} t} \rho(0) \ .
\end{equation}
Equation (\ref{eq:C}) yields $ {\hat P}{\cal L}_{\rm tot}{\hat
P}\rho =-i {\hat P} [ H_{\rm tot}, {\hat P}\rho] =-i {\hat P} [
({\hat P}H_{\rm tot}), \rho] = -i{\hat P} [ H'_S \otimes
\openone_B + \openone_S \otimes H_B, \rho] $ , whence
\begin{eqnarray}
{\hat P} e^{{\hat P}{\cal L}_{\rm tot}{\hat P} t} \rho(0) &=&
{\hat P} e^{{\cal L}_{\rm tot}' t} \rho(0)
\nonumber \\
&=& {\hat P} \left( e^{-i H'_{\rm tot} t} \rho(0)
e^{i H'_{\rm tot} t} \right) \nonumber \\
&=& {\hat P} \left( e^{-i H'_S t} \otimes e^{-i H_B t} \rho(0)
e^{i H'_S t} \otimes e^{i H_B t} \right) \ ,
\label{QZcontrolLL}
\end{eqnarray}
where the controlled Hamiltonian $H'_{\rm tot}$ and Liouvillian
$\cL'_{\rm tot}$ are given by
\begin{eqnarray}
H'_{\rm tot} & \equiv & {\hat P} H_{\rm tot}= \sum_n
\left(P_n\otimes \openone _B\right) H_{\rm tot} \left(P_n\otimes
\openone _B\right)=H'_S\otimes\openone_B+\openone_S\otimes H_B
 \ , \label{QZcontrolH} \\
\cL'_{\rm tot} & = & {\hat P}{\cal L}_{\rm tot}{\hat P} = {\hat
P}{\cal L}_S{\hat P} + \cL_B{\hat P}= \cL'_S + \cL_B{\hat P} \ .
\label{QZcontrolL}
\end{eqnarray}
with $H'_{S} = {\hat P} H_{S}= \sum_n P_n H_{S} P_n$. Hence, as a
result of infinitely frequent measurements, the system-reservoir
coupling is eliminated and, thus, decoherence is halted. We notice
the formation of invariant \emph{Zeno subspaces} \cite{FPsuper}:
in the limit of very frequent measurements, the evolution is given
by (\ref{QZcontrolH})-(\ref{QZcontrolL}) and transitions among
different sectors of the Hilbert space vanish, yielding a
superselection rule. The subspaces are defined by the
superoperator $\hat P$ defining the measurement. The
``decoherence-free" subspace is one of these Zeno subspaces.

We will assume for simplicity that ${\hat P}$ commutes with the
system Liouvillian
\begin{equation}
\label{eq:nozeno}
{\hat P}\cL_S=\cL_S {\hat P} ,
\end{equation}
i.e.\ $H'_S={\hat P} H_S = H_S$, because our purpose is to control
decoherence and we are not interested in a QZE over the system
Hamiltonian $H_S$. The above assumption is equivalent to the
following hypothesis on the Hamiltonian
\begin{equation}
\label{eq:nozeno1}
\left[ P_n,H_S\right]=0, \qquad \forall n \ .
\end{equation}
In such a case
\begin{equation}\label{eq:QZcontrolH1}
 \cL'_{\rm tot} =  (\cL_S + \cL_B){\hat P} \ .
\end{equation}

\subsection{Control via Quantum Dynamical Decoupling and ``Bang-Bang" Pulses}

We now turn our attention to the so-called quantum dynamical
decoupling \cite{BBDD,Zanardi,Vitali}, of which ``bang-bang"
pulses can be viewed as a particular case. The control of
decoherence is achieved via a time-dependent {\it system}
Hamiltonian $H_c(t)$:
\begin{equation}
H(t) =H_{\rm tot} + H_c(t) \otimes \openone_B \ ,
\label{bb-Hamiltonian}
\end{equation}
where $H_c(t)$ is designed so that $U_c(t) \equiv {\cal T}
\exp\left( -i \int_0^t H_c(s) ds\right)$ (${\cal T}$ denotes time
ordering) satisfies
\begin{eqnarray}
& & U_c(t+\tau) =U_c(t)
\label{eq:A}\\
& & \int_0^{\tau} dt \bigl( U_c^\dag(t) \otimes \openone _B\bigr)
H_{SB} \bigl( U_c(t) \otimes \openone_B \bigr) =0. \label{eq:B}
\end{eqnarray}
In the interaction picture in which $H_c(t)$ is unperturbed, the
density matrix at time $t=N \tau$, with initial state $\rho(0)$,
is given by $\rho(t) = U_{\rm tot}(N\tau) \rho(0) U_{\rm tot}^\dag
(N\tau)$ where
\begin{equation}
U_{\rm tot}(N\tau) = {\cal T} \exp\left( -i \int_0^{N\tau}
{\widetilde H}_{\rm tot}(s) ds\right) = \left[ {\cal T} \exp\left(
-i \int_0^{\tau} {\widetilde H}_{\rm tot}(s) ds\right)\right]^N
\end{equation}
and ${\widetilde H}_{\rm tot}(t)= \bigl(U_c^\dag(t) \otimes
\openone_B\bigr)
 H_{\rm tot} \bigl(U_c(t) \otimes \openone_B\bigr)$.
The second equality follows from the periodicity of ${\widetilde
H}_{\rm tot}(t)$. A standard Magnus expansion of the time-ordered
exponential \cite{Magnus} leads to
\begin{equation}
 {\cal T} \exp\left(
-i \int_0^{\tau} {\widetilde H}_{\rm tot}(s) ds\right) =
e^{-i[{\bar H}^{(0)}+{\bar H}^{(1)}+\cdots]\tau}
\end{equation}
where ${\bar H}^{(0)}\equiv {1\over \tau} \int_0^{\tau} {\tilde
H}_{\rm tot}(s) ds$ and the term ${\bar H}^{(j)}$ is of order
$\tau^j$ ($j=1,2,\cdots$). By assumption (\ref{eq:B}), one has
\begin{equation}
{\bar H}^{(0)} = H'_S \otimes \openone_B + \openone_S \otimes H_B=
H'_{\rm tot} \ ,
\end{equation}
which is formally identical to (\ref{QZcontrolH}), where $H'_S
\equiv {1\over \tau} \int_0^{\tau} dt U_c^\dag(t) H_{S}
U_c(t)=\int_0^{1} dx U_c^\dag(x\tau) H_{S} U_c(x\tau)$ is
independent of $\tau$ because $U_c(t)$ is $\tau$-periodic by
(\ref{eq:A}) and is always written as a function of $t/\tau$:
$U_c(t)=V(t/\tau)$. Therefore, in the limit $\tau\to 0$ while
keeping $t=N \tau$ constant, one obtains
\begin{equation}
\label{UnitaryBanglim}
U_{\rm tot}(t) = \left[ 1- i H'_{\rm tot} \tau + {\rm
O}\left(\tau^2\right) \right]^{t\over\tau} \stackrel{\tau \to
0}{\longrightarrow} e^{-i H'_{\rm tot} t} = e^{-i H'_S t} \otimes
e^{-i H_B t} \ .
\end{equation}
In short, as a result of the infinitely fast control, the
system-reservoir coupling is eliminated and, thus, decoherence is
halted. As we shall see in a while, this is a consequence of the
formation of invariant subspaces.

As is well known, dynamical decoupling is a generalization of the
evolution obtained by acting on the system with ``bang-bang"
pulses \cite{BBDD}. In the latter, particular case, one applies
during a time interval $\tau$ two \emph{instantaneous} unitary
operators $U_{\mathrm{k}}$ and $U_{\mathrm{k}}^\dagger$ and gets
\cite{bang}
\begin{equation}
\label{UnitaryBangH} H'_{\rm tot}= {\hat P} H_{\rm tot}= \sum_n
\left(P_n\otimes \openone _B\right) H_{\rm tot} \left(P_n\otimes
\openone _B\right)
\end{equation}
[see (\ref{QZcontrolH})], where the projections $P_n$ arise from
the spectral decomposition
\begin{equation}
U_{\mathrm{k}}=\sum_n e^{-i\lambda_n} P_n, \qquad
(\lambda_n\neq\lambda_m \; \mathrm{mod}\; 2\pi, \quad \mbox{for}
\; n\neq m) \ .
\label{eq:specdec}
\end{equation}
Notice that the map ${\hat P}$ is in this case the projection onto
the commutant
\andy{eq:centro}
\begin{equation}
\label{eq:centro}
Z(U_{\mathrm{k}})=\{X|\; [X, U_{\mathrm{k}}]=0\}.
\end{equation}
Equation (\ref{UnitaryBangH}) yields a convenient explicit
expression of the effective Hamiltonian. As in the case discussed
in the previous subsection, one observes the formation of
invariant Zeno subspaces: transitions among different subspaces
vanish in the $\tau \to 0$ limit, yielding a superselection rule.
In this case, the subspaces are defined by
(\ref{UnitaryBangH})-(\ref{eq:specdec}) and are nothing but the
ergodic sectors of $U_{\mathrm{k}}$.

By assuming again, as in (\ref{eq:nozeno})-(\ref{eq:nozeno1}),
that ${\hat P} H_S = H_S$ and that ${\hat P} H_{SB}=0$, as in
(\ref{eq:C})-(\ref{eq:C1}), we get that the controlled evolution
for $\tau\to 0$ is given by
\begin{equation}\label{eq:Utotcont}
U_{\rm tot}(t)= e^{-i H'_{\rm tot} t}=e^{-i H_S t} \otimes e^{-i
H_B t}
\end{equation}
or, in terms of Liouvillians, by $e^{\cL'_{\rm tot} t}$ with
$\cL'_{\rm tot}={\hat P}\cL_{\rm tot} {\hat P}=(\cL_S+\cL_B){\hat
P}$, exactly as in (\ref{eq:QZcontrolH1}).

Moreover, in Ref.\ \cite{bang} it was shown that one can obtain
the same result (\ref{UnitaryBangH}) by repeating a single
``bang", i.e.\ by using a single instantaneous unitary operator
$U_{\mathrm{k}}$, \emph{without} closing the group with
$U_{\mathrm{k}}^\dagger$. For simplicity, in the following we will
always consider such a situation and will assume the commutation
property (\ref{eq:nozeno}). In such a case, the evolution is
conveniently expressed in terms of the Liouvillian and density
matrix
\begin{equation}
\rho(t) = \left[e^{\cL_\mathrm{k}} e^{{\cal L}_{\rm tot} \tau}
\right]^{\frac{t}{\tau}}  {\hat P} \rho(0) \to
e^{\cL'_{\mathrm{tot}}t} {\hat P} \rho(0), \qquad \tau\to 0 \ ,
\label{eq:limcLk}
\end{equation}
where $\cL_\mathrm{k}$ is the Liouvillian corresponding to the
evolution (\ref{eq:specdec}) and $\cL'_{\rm tot}$ is given by
(\ref{eq:QZcontrolH1}). Note that the controlled Hamiltonians for
bang bang pulses, (\ref{UnitaryBangH}), and for the Zeno control,
(\ref{QZcontrolH}), coincide when the set of orthogonal
projections (\ref{measurement}) is chosen equal to the set
(\ref{eq:specdec}) of eigenprojections of $U_{\rm k}$, namely
\begin{equation}\label{eq:hatPcLk}
\cL_\mathrm{k} {\hat P} = 0, \qquad  ({\hat P}\openone)=\openone.
\end{equation}
Therefore, the two controls are equivalent in the ideal (limiting)
case \cite{bang}. However, throughout this article, by dynamical
decoupling we will refer to a situation where the evolution is
coherent (unitary), while by Zeno control to a situation where the
evolution involves incoherent (nonunitary) processes, such as
quantum measurements.

The index ``$\mathrm{k}$" in the above expressions stands for
``kicks." In the following, we shall use the expressions
``bang-bang" pulses and ``kicks" interchangeably. The latter is
reminiscent of quantum chaos \cite{qch1}. In fact, there is an
interesting link between quantum chaotic dynamics, quantum
diffusion processes and (inverse) quantum Zeno effect \cite{qch2}.
We will not elaborate on this issue in the present article.

\subsection{Control via a strong continuous coupling}

The formulation in the preceding sections hinges upon
instantaneous processes, that can be unitary or nonunitary.
However, as explained in the Introduction, the basic features of
the QZE can be obtained by making use of a continuous coupling,
when the external system takes a sort of steady ``gaze" at the
system of interest. The mathematical formulation of this idea is
contained in a theorem \cite{FPsuper} on the (large-$K$) dynamical
evolution governed by a \emph{generic} Hamiltonian of the type
\andy{HKcoup} \beq H_K= H_{\mathrm{tot}} + K
H_{\mathrm{c}}\otimes\openone_B ,
 \label{eq:HKcoup}
\eeq
which again need not describe a \textit{bona fide} measurement
process: $H_{\mathrm{c}}$ can be viewed as an ``additional"
interaction Hamiltonian performing the ``measurement" and $K$ is a
coupling constant.

Consider the time evolution operator
\barr
U_{K}(t) = \exp(-iH_K t) .
\label{eq:measinter}
\earr
In the infinitely strong measurement (infinitely quick detector)
limit $K\to\infty$, the dominant contribution is $\exp(-i K
H_{\mathrm{c}} t)$. One therefore considers the limiting evolution
operator
\beq
\label{eq:limevol}
{\mathcal{U}}(t)=\lim_{K\to\infty}\exp(i K
H_{\mathrm{c}} t)\,U_{K}(t),
\eeq
that can be shown to have the
form
\beq
\label{eq:theorem}
{\mathcal{U}}(t)=\exp(-i H'_{\mathrm{tot}} t), \eeq where \beq
H'_{\mathrm{tot}}=\hat P H_{\mathrm{tot}} =\sum_n (P_n\otimes
\openone_B) H_{\mathrm{tot}} (P_n\otimes \openone_B) \ ,
\label{eq:diagsys}
\eeq
$P_n$ being the eigenprojection of $H_{\mathrm{c}}$ belonging to
the eigenvalue $\eta_n$
\beq
\label{eq:diagevol}
H_{\mathrm{c}} = \sum_n \eta_n P_n, \qquad
(\eta_n\neq\eta_m, \quad \mbox{for} \; n\neq m) \ .
\eeq
By designing $H_c$ so that $\hat{P} H_{SB}=0$, the
system-reservoir coupling is eliminated and, thus, decoherence is
halted. Equation (\ref{eq:diagsys}), restricted to the system of
interest, is formally identical to (\ref{UnitaryBangH}) and
(\ref{QZcontrolH}).

In conclusion, the limiting evolution operator is
\andy{eq:measinterbis} \barr U_K(t)\sim\exp(-i K H_{\mathrm{c}}
t)\,{\mathcal{U}}(t) = \exp\left(-i K  t\sum_n \eta_n
P_n\otimes\openone_B -i \hat{P} H_{\mathrm{tot}} t\right) .
\label{eq:measinterbis} \earr The above statements can be proved
by making use of the adiabatic theorem \cite{Messiah61}. It is
worth noting that the evolution in the strong coupling limit is
known to force the system to ``cling" to the eigenstates of the
interaction \cite{Frasca}. In this sense, one expects that the
dynamics be dominated by $H_{\mathrm{c}}$ for $K$ large. The above
theorem clarifies how the structure of $H_{\mathrm{c}}$ determines
the features of the dynamics. Once again, like in the two previous
sections, one observes the formation of invariant Zeno subspaces,
that are in this case the eigenspaces of the interaction
(\ref{eq:diagsys})-(\ref{eq:diagevol}): the block-diagonal
structure of (\ref{eq:measinterbis}) is explicit. The links
between the quantum Zeno effect and the notion of ``continuous
coupling" to an external apparatus or environment has often been
proposed in the literature of the last 25 years
\cite{Simonius,HS,Schulman98,varicont}. The novelty here lies in
the gradual formation of the Zeno subspaces as $K$ becomes
increasingly large. In such a case, they are nothing but the
adiabatic subspaces. In terms of the Liouvillian:
\begin{equation}
\rho(t) = e^{(K\cL_c+\cL_{\mathrm{tot}})t} {\hat P} \rho(0)  \to
e^{\cL'_{\mathrm{tot}}t} {\hat P} \rho(0), \qquad K\to \infty \ ,
\end{equation}
[see (\ref{eq:limcLk})] where the notation is obvious and
\begin{equation}\label{eq:hatPcLc}
\cL_c {\hat P} = 0, \qquad  ({\hat P}\openone)=\openone
\end{equation}
[see (\ref{eq:hatPcLk})]. The Liouvillian
$\cL'_{\mathrm{tot}}={\hat P}\cL_{\mathrm{tot}}{\hat P}$
corresponds to $\hat{P} H_{\mathrm{tot}} = H'_{\mathrm{tot}}$ and,
under the assumption (\ref{eq:nozeno}) and (\ref{eq:C1}), is again
given by (\ref{eq:QZcontrolH1}).

\subsection{Controlled evolution and Zeno subspaces}

\begin{figure}
\begin{center}
\includegraphics[height=5cm]{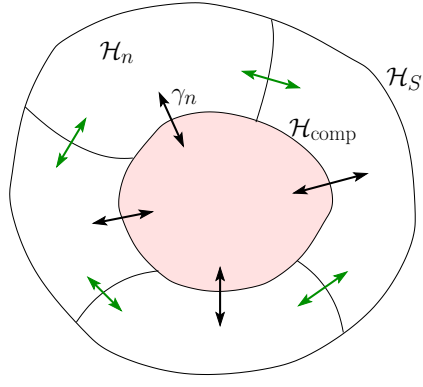}
\end{center}
\caption{The Zeno subspaces are formed when the frequency
$\tau^{-1}$ of measurements or BB pulses or the strength $K$ of
the continuous coupling tend to $\infty$. The shaded region
represents the ``computational" subspace ${\cal H}_{\rm comp}
\subset {\cal H}_S$ defined in Eq.\ (\ref{eq:directsum}). The
transition rates $\gamma_n$ depend on $\tau$ or $K$.}
\label{fig:zenosub}
\end{figure}

The three different procedures described in this section yield, by
different physical mechanisms, the formation of invariant Zeno
subspaces. This is shown in Fig.\ \ref{fig:zenosub}. If one of
these invariant subspaces is the ``computational" subspace ${\cal
H}_{\rm comp}$ introduced in Eq.\ (\ref{eq:directsum}), the
possibility arises of inhibiting decoherence in this subspace.

Of course, in the $\tau,K^{-1} \to 0$ limit, decoherence can be
completely \emph{halted}, according to Eqs.\
(\ref{QZcontrolLL})-(\ref{QZcontrolL}),
(\ref{UnitaryBanglim})-(\ref{UnitaryBangH}) and
(\ref{eq:diagsys})-(\ref{eq:diagevol}). However, the objective of
our study is to understand \emph{how} the limit is attained and
analyze the deviations from the ideal situation. This will be done
by studying the transition rates $\gamma_n$ between different
subspaces and in particular their $\tau$ and $K$ dependence (see
Fig.\ \ref{fig:zenosub}). We shall see that in general this
dependence can be complicated, leading to \emph{enhancement} of
decoherence in some cases and \emph{suppression} in other cases.
For this reason, the \emph{physical} meaning of the expressions
$\tau,K^{-1} \to 0$ in this section must be scrutinized with great
care.

\section{Master equation}
\label{sec.model}

We consider the time evolution when the initial state is
factorized as in (\ref{eq:rhoprod}) and the reservoir equilibrium
state has inverse temperature $\beta$
\begin{equation}
\label{eq:incond} \rho_B = {1\over Z} \exp(-\beta H_B)
\label{BathEq}={1\over Z}\sum_\mu \exp(-\beta
E_\mu)|\mu\ket\bra\mu| , \qquad (\cL_B \rho_B=0)
\end{equation}
where $Z=\mathrm{tr}_B e^{-\beta H_B}=\sum_\mu e^{-\beta E_\mu}$
is the normalization constant. We will assume throughout our
analysis that the characteristic timescales of quantum state
manipulation in the space ${\cal H}_{\rm comp}$ [see
(\ref{eq:directsum})] are much longer than any other timescales,
so that the process is well described by the van Hove $``\lambda^2
t"$ limit \cite{vanHove,SpohnLeb,AVL,Yuasa}, where $\lambda$ is
the coupling constant between system and reservoir [see the
comment after Eq. (\ref{Ltotdef})]. For instance, if we take the
timescale of quantum state manipulation to be of order
$\lambda^{-2}$ ($\sim$ to a Rabi period in ${\cal H}_{\rm comp}$),
then the other energies involved are at most O($\lambda^{0}$). We
will look at some concrete examples in Sec.\ \ref{sec:examples}.

Following Gardiner and Zoller \cite{GardinerZoller}, we now
quickly derive the master equation and set up our notation. The
starting point is the decomposition of the Liouville equation with
the aid of the projection operators
\begin{eqnarray}
\label{eq:defproj}
{\cal P}\rho = \mathrm{tr}_B \{\rho\} \otimes \rho_B= \sigma
\otimes \rho_B, \qquad {\mathcal{Q}}=1-{\mathcal{P}}
\end{eqnarray}
where ${\rm tr}_B$ stands for the partial trace over the reservoir
degrees of freedom and $\rho_B$ is the equilibrium reservoir state
(\ref{BathEq}). Note that ${\mathcal{P}}^2={\mathcal{P}}$ and
${\mathcal{Q}}^2={\mathcal{Q}}$. Moreover,
\begin{equation}\label{eq:propPQ}
{\mathcal{P}}\cL_S=\cL_S {\mathcal{P}}, \qquad
{\mathcal{P}}\cL_B=\cL_B {\mathcal{P}}=0
\end{equation}
and we assume that
\begin{equation}\label{eq:propPQ1}
{\mathcal{P}}\cL_{SB}{\mathcal{P}}=0,
\end{equation}
which can always be satisfied by redefining the system Liouville
operator $\cL_S \rho \to \cL_S \rho +\mathrm{tr}_B \{\cL_{SB}
\rho\}\otimes\rho_B$ and the interaction Liouville operator
$\cL_{SB}\rho \to \cL_{SB}\rho -\mathrm{tr}_B \{\cL_{SB}
\rho\}\otimes\rho_B$.

The evolution in the interaction picture reads
\begin{equation}
\dot \rho_{\mathrm{I}}(t)= \cL_{SB}(t) \rho_{\mathrm{I}}(t),
\qquad \cL_{SB}(t)= e^{-\cL_0 t} \cL_{SB} e^{\cL_0 t}
\label{eq:propnew}
\end{equation}
and by applying the projection (\ref{eq:defproj}) together with
Eq.\ (\ref{eq:propPQ1}) one gets
\begin{eqnarray}
{\mathcal{P}}\dot \rho_{\mathrm{I}}(t) = {\mathcal{P}}
\cL_{SB}(t){\mathcal{Q}}\rho_{\mathrm{I}}(t) , \qquad
{\mathcal{Q}}\dot \rho_{\mathrm{I}}(t) = {\mathcal{Q}}
\cL_{SB}(t){\mathcal{P}}\rho_{\mathrm{I}}(t)+ {\mathcal{Q}}
\cL_{SB}(t){\mathcal{Q}}\rho_{\mathrm{I}}(t).
\end{eqnarray}
By formally integrating the second equation and plugging the
result into the first one, one obtains to order $\lambda^2$
\begin{equation}
\label{eq:projeq}
 {\mathcal{P}}\dot \rho_{\mathrm{I}}(t) = \int_0^t ds\;
{\mathcal{P}}\cL_{SB}(t){\mathcal{Q}}\cL_{SB}(s)
{\mathcal{P}}\rho_{\mathrm{I}}(s),
\end{equation}
where the initial condition (\ref{eq:rhoprod}), yielding
${\mathcal{Q}}\rho_{\mathrm{I}}(0)=0$, was used. By using the
definitions (\ref{eq:defproj}) and the conditions
(\ref{eq:propPQ})-(\ref{eq:propPQ1}), Eq.\ (\ref{eq:projeq})
yields
\begin{equation}\label{eq:nonMarkov}
\dot\sigma_{\mathrm{I}}(t)=\int_0^t ds\,
{\mathcal{K}}_{\mathrm{I}}(t,s) \sigma_{\mathrm{I}}(s) ,
\end{equation}
where
\begin{equation}\label{eq:cKI}
\cK_{\mathrm{I}}(t,s)\, \sigma = \mathrm{tr}_B
\left\{\cL_{SB}(t)\cL_{SB}(s)\, \sigma\otimes\rho_B \right\}.%
\end{equation}
By making use of the first Markov approximation
$\sigma_{\mathrm{I}}(s)\to\sigma_{\mathrm{I}}(t)$
\cite{GardinerZoller}, which is motivated by the fact that the bath
correlation kernel $\cK_{\mathrm{I}}(t,s)$ is different from zero
only for $s\simeq t-\tau_c$ such that
$\sigma_{\mathrm{I}}(t-\tau_c)\simeq\sigma_{\mathrm{I}}(t)$, one
gets
\begin{equation}\label{eq:MarkovI}
\dot\sigma_{\mathrm{I}}(t)=\cL(t)
\sigma_{\mathrm{I}}(t), \qquad \cL(t)= \int_0^t ds \,
{\mathcal{K}}_{\mathrm{I}}(t,s) \ .
\end{equation}
If the time $t$ in Eqs.~(\ref{eq:MarkovI}) is much larger than the
bath correlation time, $t\gg\tau_c$, one can safely replace the
upper limit of integration with $\infty$, getting a Markovian
equation with the time independent Liouville operator
$\cL=\cL(\infty)$.

We emphasize that this procedure can be rigorously justified in
the (weak coupling) limit \cite{AVL}
\begin{equation}\label{eq:formal}
\cL= \lim_{\lambda \to 0} \int_0^{t/\lambda^2} ds \,
{\mathcal{K}}_{\mathrm{I}}(t/\lambda^2,s),
\end{equation}
which physically corresponds to a time coarse-graining ansatz
\cite{Pauli,LBW}. From (\ref{eq:cKI}) and (\ref{eq:formal}) one
gets (by suppressing, for simplicity, the subscript I for the
operators in the interaction picture)
\begin{eqnarray}
\cL\, \sigma &=& \lim_{\lambda \to 0}\;\mathrm{tr}_B
\left\{e^{-\cL_S\, t/\lambda^2} \left[\int_{-t/\lambda^2}^0 ds\;
\cL_{SB}\cL_{SB}(s)\right]e^{\cL_S\, t/\lambda^2}\,
\sigma\otimes\rho_B \right\} \nonumber \\
& =& \mathrm{tr}_B \left\{\sum_\omega \tilde Q_\omega
\left[\int_{-\infty}^0 ds\; \cL_{SB}\cL_{SB}(s)\right]\tilde
Q_\omega\, \sigma\otimes\rho_B \right\} ,
\label{eq:formal1}
\end{eqnarray}
where $\tilde Q_\omega$ are the eigenprojections of the
Liouvillian $\cL_S$,
\begin{equation}
\cL_S=-i \sum_\omega \omega\tilde Q_\omega,\quad \sum_\omega\tilde
Q_\omega=1,\quad\tilde Q_\omega\tilde Q_{\omega'}
=\delta_{\omega,\omega'}\tilde Q_\omega \ ,
\label{eq:eigenA9}
\end{equation}
and in the limit the off-diagonal terms,
$e^{i(\omega-\omega')t/\lambda^2}\tilde Q_\omega [\cdots]\tilde
Q_{\omega'}$, vanish due to the Riemann-Lesbegue lemma. Notice
that the superoperators  $\tilde Q_\omega$ can be expressed in
terms of the eigenprojections of the Hamiltonian $H_S$ as
\begin{equation}\label{eq:supord}
\tilde Q_\omega \rho = \sum_{\scriptstyle m,n \atop \scriptstyle
E_m-E_n=\omega} Q_m \rho Q_n, \qquad H_S = \sum_n E_n Q_n \ .
\end{equation}

 From a physical point of view, the result (\ref{eq:formal1}) hinges upon
a second-order perturbation expansion of the Liouvillian
(\ref{Ltotdef}) in the interaction picture
\begin{eqnarray}\label{eq.2ndorder}
 e^{-\cL_0 t} e^{{\cal L}_{\rm tot} t} =
{\cal T} \exp\left(\int_0^{t} ds\; \cL_{SB}(s)\right)
 \simeq \openone + \int_0^t ds\;\cL_{SB}(s) +
 \int_0^t ds\int_0^s ds_1\;\cL_{SB}(s)\cL_{SB}(s_1).
\end{eqnarray}
Indeed, the first-order term vanishes after the projection due to
(\ref{eq:propPQ1}), while the projected second-order term reads
\begin{eqnarray}
& &\mathrm{tr}_B \left\{ \int_0^t ds\int_0^s
ds_1\;\cL_{SB}(s)\cL_{SB}(s_1)\, \sigma\otimes\rho_B \right\}=
\int_0^t ds\;\mathrm{tr}_B \left\{e^{-\cL_S s} \left[\int_{-s}^0
ds_1\; \cL_{SB}\cL_{SB}(s_1)\right]e^{\cL_S s}\,
\sigma\otimes\rho_B \right\} \nonumber\\ & &\simeq \int_0^t
ds\;\mathrm{tr}_B\left\{e^{-\cL_S s} \left[\int_{-\infty}^0 ds_1\;
\cL_{SB}\cL_{SB}(s_1)\right]e^{\cL_S s}\, \sigma\otimes\rho_B
\right\}\simeq t\;\mathrm{tr}_B \left\{\sum_\omega \tilde Q_\omega
\left[\int_{-\infty}^0 ds\; \cL_{SB}\cL_{SB}(s)\right]\tilde
Q_\omega\, \sigma\otimes\rho_B \right\} \nonumber\\
& & = \cL t\;\sigma \ .
\label{eq:P2ndorder}
\end{eqnarray}
In the second equality we considered times $t$ much larger than
the bath correlation time $\tau_c$, so that the integration range
can be extended from $(-s, 0)$ to $(-\infty, 0)$, while in the
third equality we neglected the rapidly oscillating (compared with
those responsible for decoherence) off-diagonal terms. By
combining (\ref{eq:P2ndorder}) and (\ref{eq.2ndorder}) we finally
get
\begin{equation}\label{eq:2ndfinal}
\sigma_{\mathrm{I}}(t)= \mathrm{tr}_B \left\{e^{-\cL_0 t} e^{{\cal
L}_{\rm tot} t}\, \sigma_{\mathrm{I}}(0)\otimes\rho_B
\right\}\simeq \exp\left(\cL t\right) \sigma_{\mathrm{I}}(0) \ ,
\end{equation}
which is nothing but (\ref{eq:MarkovI}), when one substitutes
$\cL(t)\to\cL(\infty)=\cL$.

Some of these ideas and techniques, at different levels of rigor,
have been investigated and applied in the literature of the last
four decades \cite{vanHove,SpohnLeb,Yuasa}.

\subsection{The general case}
\label{sec.modelA}

Assume now that the interaction Hamiltonian $H_{SB}$ can be
written as \cite{GardinerZoller}
\begin{equation}\label{eq:intHam}
H_{SB}=\sum_m \left(X_m \otimes A^\dagger_m+ X^\dagger_m\otimes
A_m\right),
\end{equation}
where the $X_m$ are the eigenoperators of the system Liouvillian,
satisfying
\begin{equation}\label{eq:eigenoperators}
\cL_S X_m= i \omega_m X_m \qquad (\omega_m\neq\omega_n, \quad
\mathrm{for}\quad m\neq n)
\end{equation}
and $A_m$ are destruction operators of the bath
\begin{equation}\label{eq:Am}
 A_m=A(g_m)=\int d^3 k\; g_m^*(\bm k)\; a(\bm k) \ ,
\end{equation}
expressed in terms of bosonic operators $a(\bm k)$, with form
factors $g_m(\bm k)$. We are specifying our analysis to three
dimensions (although it is valid in any dimensions). Incidentally,
the form of the Hamiltonian (\ref{eq:intHam}) is of very general
validity (and is not limited, as one might naively think, to
dipole-like approximations): the only assumption made is that the
coupling with the bath be linear, i.e.\ one is not considering
terms of the type $a^2, a^{\dagger 2}$, etc., which would only be
relevant for squeezed reservoirs. In practice, one determines the
operators (\ref{eq:eigenoperators}), then finds the bath operators
in order to write the interaction in the form (\ref{eq:intHam}),
and neglects nonlinear terms.

In the interaction representation we get
\begin{equation}\label{eq:intHamint}
H_{SB}(t)=e^{-\cL_0 t}H_{SB}=\sum_m \left(X_m \otimes
A^\dagger_m(t)+ X^\dagger_m\otimes A_m(t)\right),
\end{equation}
where
\begin{equation}\label{eq:Amt}
 A_m(t)=\int d^3 k\; g_m^*(\bm k)\;  e^{-i(\omega_k-\omega_m)t} a(\bm k) \ .
\end{equation}
If the bath is in the thermal state (\ref{BathEq}) we obtain
\begin{eqnarray}
\left\langle A_m(t) A_m^\dagger(s) \right\rangle &=& \int d^3  k\;
|g_m(\bm k)|^2 \left(N(\omega_k)+1\right) e^{-i(\omega_k-\omega_m)
(t-s)}   \ , \nonumber\\
\left\langle A_m^\dagger(t) A_m(s) \right\rangle &=& \int d^3  k\;
|g_m(\bm k)|^2 N(\omega_k) e^{i(\omega_k-\omega_m) (t-s)} \,
\label{eq:corr}
\end{eqnarray}
and $\left\langle A_m(t) A_m(s) \right\rangle=\left\langle
A_m^\dagger(t) A_m^\dagger(s) \right\rangle=0$, with $N(\omega)=
1/(e^{\beta\omega}-1)$.

From (\ref{eq:formal1}) we get
\begin{equation}\label{eq:cLf}
\cL \sigma = \int_{-\infty}^0 ds\;\mathrm{tr}_B \left\{\tilde
Q_\omega \cL_{SB}\cL_{SB}(s)\tilde Q_{\omega}\sigma\otimes\rho_B
\right\}
\end{equation}
and by using the property
\begin{equation}\label{eq:prl1l2}
\sum_\omega \tilde Q_\omega \cL_1 \cL_2 \tilde Q_\omega \rho=-
\sum_\omega \tilde Q_\omega \left[H_1,\left[H_2, \tilde Q_\omega
\rho\right]\right]=-\sum_\omega \left[(\tilde Q_\omega H_1),
\left[(\tilde Q_{-\omega} H_2),\rho\right]\right] \ ,
\end{equation}
which easily follows from the definition (\ref{eq:eigenA9}), we
get
\begin{equation}\label{eq:diagpwa1cKI34}
\cL \sigma = -\sum_\omega \int_{-\infty}^0 ds\; \mathrm{tr}_B
\left\{\left[\left(\tilde Q_\omega
H_{SB}\right),\left[\left(\tilde Q_{-\omega}
H_{SB}(s)\right),\sigma\otimes\rho_B\right]\right] \right\} .
\end{equation}
By using (\ref{eq:supord}) and (\ref{eq:intHam}) one obtains
\begin{equation}\label{eq:pwHsb}
\tilde Q_{\omega_m} H_{SB}= H_{SB}^{(m)}=X_{-m} \otimes
A^\dagger_{-m} + X^\dagger_m \otimes A_m,
\end{equation}
whence
\begin{eqnarray}
\cL\sigma &=& -\sum_m \int_{-\infty}^0 ds\; \mathrm{tr}_B \left\{
\left[H_{SB}^{(m)},\left[ H_{SB}^{(-m)}(s),\sigma\otimes\rho_B\right]\right]
\right\} \nonumber\\
& = & -\sum_m \int_{-\infty}^0 ds\; \mathrm{tr}_B \left\{
\left[X^\dagger_m \otimes A_m,\left[ X_{m} \otimes
A^\dagger_{m}(s),\sigma\otimes\rho_B\right]\right]
 + \left[X_{-m}
\otimes A^\dagger_{-m},\left[ X^\dagger_{-m} \otimes
A_{-m}(s),\sigma\otimes\rho_B\right]\right]\right\} \ . \
\label{eq:diagcL}
\end{eqnarray}
In the second equality we neglected terms containing two
annihilation or creation operators, which identically vanish by
performing the trace over the thermal state $\rho_B$. Equation
(\ref{eq:diagcL}) can be put in the form \cite{GardinerZoller}
\begin{eqnarray}
\cL \sigma &=& -i \sum_m \left[\delta_m X_m^\dagger X_m +
\epsilon_m X_m X_m^\dagger, \sigma\right] \nonumber\\
& & + \sum_m K_m\left(X_m\sigma
X_m^\dagger-\frac{1}{2}\left\{X_m^\dagger X_m,
\sigma\right\}\right)+ \sum_m G_m\left(X_m^\dagger\sigma
X_m-\frac{1}{2}\left\{X_m X_m^\dagger, \sigma\right\}\right) \ ,
\label{eq:Liouvillian}
\end{eqnarray}
and
\begin{eqnarray}
  \frac{1}{2} K_m-i\delta_m = \int_0^\infty dt
  \left\langle A_m(0) A_m^\dagger(t) \right\rangle \ , \qquad
  \frac{1}{2} G_m-i\epsilon_m  = \int_0^\infty dt
  \left\langle A_m^\dagger(0) A_m(t) \right\rangle \ .
  \label{eq:renormfactors}
\end{eqnarray}
The first line in (\ref{eq:Liouvillian}) is just the
renormalization of the free Liouvillian $\cL_S$ by Lamb and Stark
shift terms and will be neglected in the following. The
dissipative part is given by the second line, which appears in the
Lindblad form, so that $\mathrm{tr \cL\sigma}=0$.

In Eq.\ (\ref{eq:eigenoperators}) we will identify
$\omega_{-m}=-\omega_m$, and will assume that $X_{-m}=X_m^\dagger$
and $g_m=g_{-m}$, which is equivalent to the hypothesis that the
interaction Hamiltonian be the product of selfadjoint operators
acting on the system and the bath, namely $H_{SB}=\sum_i
H_{S}^{(i)}\otimes H_{B}^{(i)}$, with $H_{S}^{(i)}$ and
$H_{B}^{(i)}$ selfadjoint. Notice, therefore, that we are
\emph{not} making any rotating-wave approximation, and the
interaction Hamiltonian $H_{SB}$ (\ref{eq:intHam}) contains
\emph{both} rotating and counter-rotating terms. The dissipative
part of (\ref{eq:Liouvillian}) can now be rewritten as
\begin{eqnarray}
\cL \sigma &=&  \gamma_0\left(X_0\sigma X_0-\frac{1}{2}\left\{X_0
X_0, \sigma\right\}\right)\nonumber\\
& & + \sum_{m\ge1} \gamma_m\left(X_m\sigma
X_m^\dagger-\frac{1}{2}\left\{X_m^\dagger X_m,
\sigma\right\}\right)+ \sum_{m\ge1} \gamma_{-m}
\left(X_m^\dagger\sigma X_m-\frac{1}{2}\left\{X_m X_m^\dagger,
\sigma\right\}\right) \ ,
\label{eq:Liouvilliannrw}
\end{eqnarray}
where $\gamma_m=K_m+G_{-m}$. We introduce the bare spectral
density functions (form factors)
\begin{equation}\label{eq:formfactor}
\kappa_m(\omega)=\int d^3  k\; |g_m(\bm k)|^2
\delta(\omega_k-\omega), \qquad \kappa_m(\omega)=0, \quad
\mathrm{for}\quad \omega<0
\end{equation}
and the thermal spectral density functions,
\begin{equation}\label{eq:formfactors}
\kappa_m^{\beta}(\omega)=\kappa_m(\omega)\left(N(\omega)+1\right)+\kappa_m(-\omega)
N(-\omega)=\frac{1}{1-e^{-\beta\omega}}\left[\kappa_m(\omega)-\kappa_m(-\omega)\right]
,
\end{equation}
which extend along the whole real axis due to the counter-rotating
terms and satisfy the KMS symmetry \cite{SpohnLeb}
\begin{equation}\label{eq:excsym}
\kappa_m^\beta(-\omega)=\frac{N(\omega)}{
N(\omega)+1}\,\kappa_m^\beta(\omega)= \frac{ N(-\omega)+1}{
N(-\omega)}\,\kappa_m^\beta(\omega)= \exp(-\beta \omega)\,
\kappa_m^\beta(\omega) .
\end{equation}
We explicitly get
\begin{eqnarray}
K_m= 2\pi \kappa_m(\omega_m) \, \left(N(\omega_m)+1\right), \qquad
G_m= 2\pi \kappa_m(\omega_m)\, N(\omega_m)
\label{eq:KmGmdef}
\end{eqnarray}
and
\begin{eqnarray}
\gamma_m= 2 \Re \int_0^\infty dt \left(\left\langle A_m
A^\dagger_m(t)\right\rangle+ \left\langle A_{-m}^\dagger
A_{-m}(t)\right\rangle\right) = 2\pi \kappa_m^\beta(\omega_m) \ .
\label{eq:gammamdef}
\end{eqnarray}

It is useful to look at some concrete examples and scrutinize the
modification of the form factor (\ref{eq:formfactor}), due to the
presence of the thermal bath. Let us focus, for the sake of
clarity, on two particular Ohmic cases: an exponential form factor
\beq
\label{eq:expformfactor}
\kappa_m^{(E)}(\omega)= g^2\omega \exp (-\omega/\Lambda) \theta
(\omega)
\eeq
and a polynomial form factor
\beq
\label{eq:lorformfactor1}
\kappa^{(P)}_m(\omega)= g^2\frac{\omega}{[1+(\omega/\Lambda)^2]^n}
\theta (\omega) .
\eeq
In the latter case, we focus on $n=2$, which is typical of quantum
dots \cite{qdot} (the case $n=4$ is also of interest, being the
nonrelativistic form factor of the 2P-1S transition of the
hydrogen atom \cite{Hff,PLA98}). In the above formulas, $g$ is a
coupling constant, $\Lambda$ a cutoff and $\theta$ the unit step
function.
\begin{figure}
\begin{center}
\includegraphics[width=\figwidth]{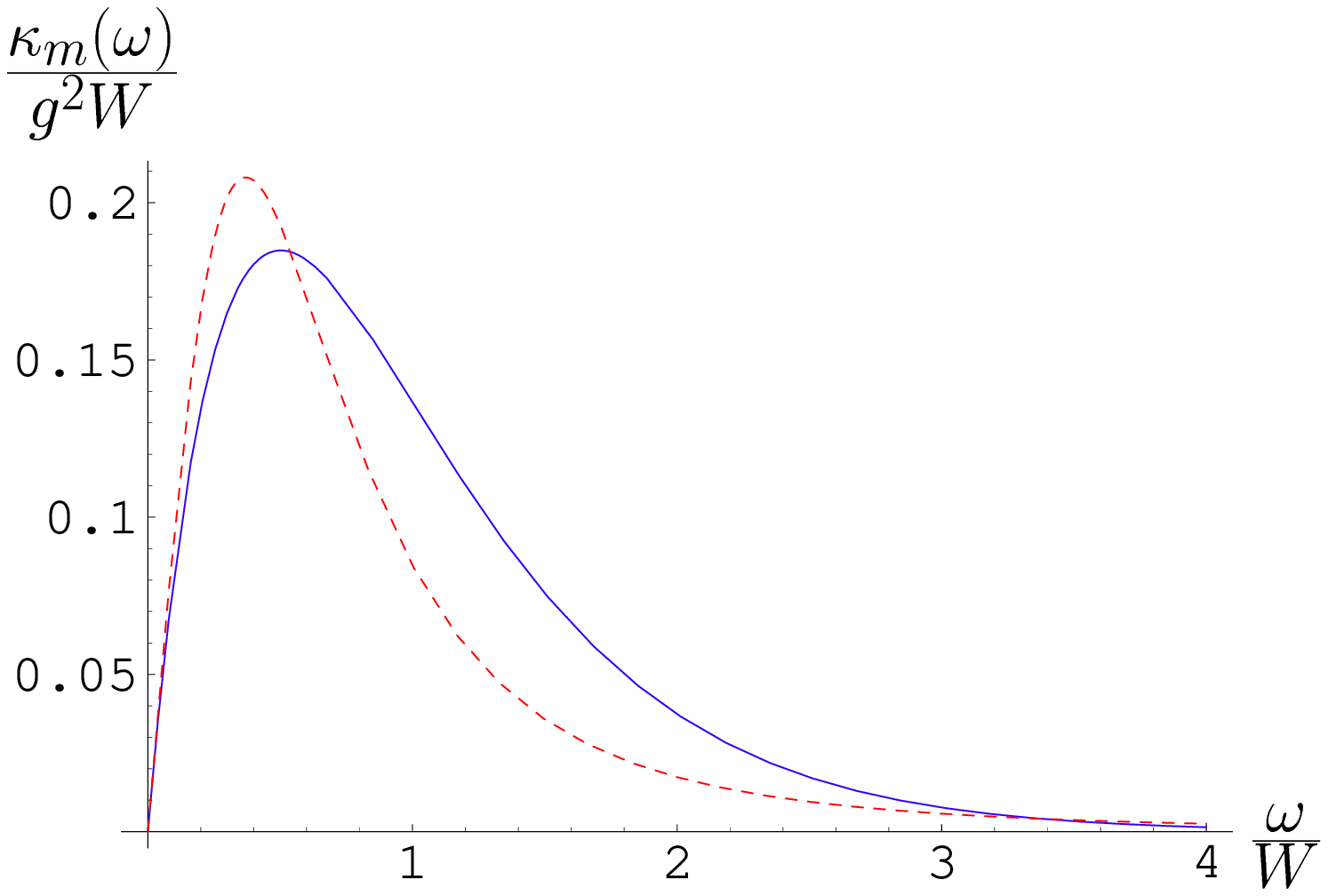}
\end{center}
\caption{The form factors at zero temperature,
$\kappa(\omega)/(g^2W)$ vs $\omega/W$. Full line: exponential form
factor (\ref{eq:expformfactor}); dashed line: polynomial form
factor (\ref{eq:lorformfactor1}). }
\label{fig:thermalzero}
\end{figure}
In order to properly compare these two cases, we will require that
the bandwidth be the same: \beq \label{eq:BW}
W={\int_{-\infty}^\infty d \omega \;
|\omega|\kappa^{(E)}_m(\omega) \over \int_{-\infty}^\infty d
\omega \;\kappa^{(E)}_m(\omega) } = { \int_{-\infty}^\infty d
\omega \;|\omega|\kappa^{(P)}_m(\omega) \over
\int_{-\infty}^\infty d \omega \;\kappa^{(P)}_m(\omega) }, \eeq
where the square root of the denominator \beq \label{eq:Ztime}
\left[\int_{-\infty}^\infty d \omega \;
\kappa_m(\omega)\right]^{-{1\over2}}\equiv\tau_{\mathrm{Z}} \eeq
is the so-called Zeno time, characterizing the convexity of the
survival probability at the origin \cite{PLA98,Antoniou,PIO}.
Notice that a \emph{finite} natural cutoff $\Lambda \simeq
8.498\cdot 10^{18} \mbox{rad/s}$ and a \emph{finite} Zeno time
$\tau_{\rm Z} \simeq 3.593 \cdot 10^{-15} \mbox{s}$ can also be
computed for the hydrogen atom in vacuum [polynomial form factor
(\ref{eq:lorformfactor1}) with $n=4$], as well as for atomic and
molecular systems whose electronic wave functions are known. The
condition (\ref{eq:BW}) when $n=2$ yields the ratio $\Lambda_{\rm
pol}/\Lambda_{\rm exp}=1.275$ between the cutoffs for the
polynomial and exponential form factors, and $W=1.99\Lambda_{\rm
exp}$. The two form factors are displayed in Fig.\
\ref{fig:thermalzero} for $g=1$.

\begin{figure}
\begin{center}
\includegraphics[width=\figwidth]{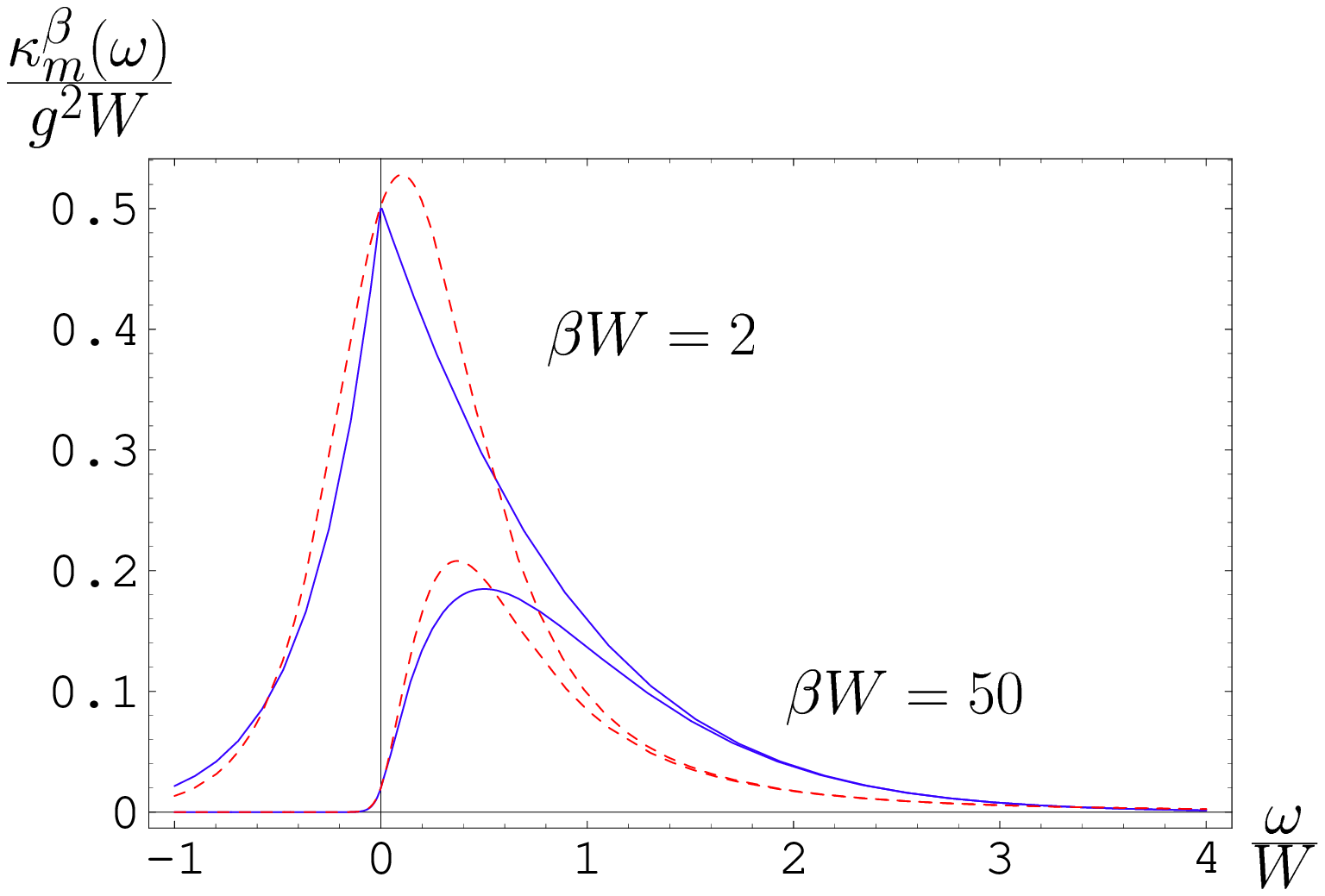}
\end{center}
\caption{The thermal form factors
$\kappa_m^\beta(\omega)$ vs $\omega$. Full lines: exponential form
factors (\ref{eq:formfactors}), (\ref{eq:expformfactor}); dashed
lines: polynomial form factors (\ref{eq:formfactors}),
(\ref{eq:lorformfactor1}). The form factor is larger at higher
temperature $\beta^{-1}$. Note the discontinuity of the derivative
in the exponential case at $\omega=0$ (more apparent at high
temperature). We set $\beta W=50$ (low temperature) and $\beta
W=2$ (very high temperature).}
\label{fig:thermal}
\end{figure}

The thermal form factors (\ref{eq:formfactors}) are displayed in
Fig.\ \ref{fig:thermal} for two different temperatures. Three
features are apparent: the form factor is an increasing function
of the temperature $\beta^{-1}$. Its value at $\omega=0$ is
$\kappa_m^{\beta}(0)= \kappa'_m(0^+)/\beta=g^2/\beta$, where the
prime denotes derivative. Moreover, its derivative reads
$\kappa_m^{\beta\prime}(0^\pm)= \kappa'_m(0^+)/2\pm
\kappa''_m(0^+)/(2\beta)$, whence it is continuous,
$\kappa_m^{\beta\prime}(0)=g^2/2$, in the polynomial case (because
$\kappa''_m(0^+)=0$),  and discontinuous,
$\kappa_m^{\beta\prime}(0^\pm)=g^2/2\mp g^2/(\beta\Lambda)$, in
the exponential case; this is more apparent at higher
temperatures. Finally, the support of the thermal form factors is
no longer lower bounded, due to the effect of the counter-rotating
terms.

\subsection{Two-level system}
\label{Tlev}

A particular case of the above is the qubit Hamiltonian
\begin{equation}\label{eq:2Hamiltonian}
H_0= \frac{\Omega}{2} \sigma_z, \qquad
H_{SB}=\sigma_z\otimes\left(A(g_0)+A^\dagger(g_0)\right)+\sigma_x\otimes\left(
A(g_1)+A^\dagger(g_1)\right) .
\end{equation}
This is of the form (\ref{eq:intHam}), when one identifies
\begin{equation}\label{eq:X+-}
X_0=\sigma_z, \quad X_{\pm 1}=\sigma_{\mp}=\frac{\sigma_x\mp
i\sigma_y}{2}, \quad \omega_{\pm 1}=\pm \Omega, \quad \omega_0=0,
\end{equation}
hence
\begin{eqnarray}
\cL \rho =  \gamma_0\left(\sigma_z\rho \sigma_z-\rho\right) +
\gamma_{+1}\left(\sigma_- \rho \sigma_+
-\frac{1}{2}\left\{\sigma_+ \sigma_-, \rho\right\}\right)+
\gamma_{-1}\left(\sigma_+ \rho \sigma_-
-\frac{1}{2}\left\{\sigma_- \sigma_+, \rho\right\}\right) \ ,
\label{eq:Liouvillian2l}
\end{eqnarray}
with
\begin{equation}\label{eq:gamma2l}
\gamma_0=2\pi \kappa_0^\beta (0)=2\pi \frac{\kappa'_0(0^+)}{\beta},
\qquad \gamma_{\pm1}=2\pi \kappa_1^\beta (\pm\Omega) \ ,
\end{equation}
where we used (\ref{eq:formfactors}).

\section{Quantum Zeno Control}
\label{sec.Zeno}

Let us look at the quantum Zeno dynamics with a \emph{finite}
interval $\tau=t/N$ between measurements,
\begin{equation}
\rho(t) = \left[{\hat P} e^{{\cal L}_{\rm tot} \tau} {\hat
P}\right]^{\frac{t}{\tau}}   \rho(0) \ ,
\end{equation}
where $\cL_{\mathrm{tot}}$ and ${\hat P}$ are given by
(\ref{Ltotdef}) and (\ref{measurement}), respectively. We will
look at the subtle effects on the decay rate arising from the
presence of the short-time quadratic (Zeno) region. Therefore the
standard method \cite{vanHove} is not applicable to the present
situation and the limit must be evaluated by a different
technique. We only sketch the main steps in the derivation and
give more details in Appendix \ref{sec:B}. Second order
perturbation in $\cL_{SB}$ and the conditions
(\ref{eq:C})-(\ref{eq:C1}) yield
\begin{eqnarray}
{\hat P} e^{{\cal L}_{\rm tot} \tau} {\hat P} = {\hat P} e^{\cL_0
\tau} {\cal T} \exp\left(\int_0^{\tau} ds\; \cL_{SB}(s)\right)
{\hat P} \simeq  e^{\cL_0 \tau} {\hat P}\left[ \openone +
\int_0^\tau ds\; \cL_{SB}(s) + \int_0^\tau ds\int_0^s ds_1\;
\cL_{SB}(s)\cL_{SB}(s_1) \right]{\hat P} \ .
\end{eqnarray}
In terms of the operator $\cG_{\mathrm{Z}}(\tau)$, defined as the
solution of the operator equation
\begin{eqnarray}
\int_0^{\tau} ds\; e^{-\cL_0 s} \cG_{\mathrm{Z}}(\tau) e^{\cL_0
s}= {\hat P}\int_0^\tau ds\int_0^s ds_1\; \cL_{SB}(s)\cL_{SB}(s_1)
{\hat P} =\int_0^{\tau} ds\; e^{-\cL_0 s}\left[\int_0^s ds_1\;
{\hat P}\cL_{SB} \cL_{SB}(-s_1){\hat P}\right] e^{\cL_0 s}
 \ ,
\end{eqnarray}
one obtains
\begin{eqnarray}
\left[{\hat P} e^{\cL_{\rm tot} \tau} {\hat P}\right]^{t/\tau}
\simeq \left[{\hat P} e^{\cL_0 \tau} {\cal T}
\exp\left(\int_0^{\tau} ds\; e^{-\cL_0 s} \cG_{\mathrm{Z}}(\tau)
e^{\cL_0 s}\right){\hat P}\right]^{t/\tau}= {\hat P}
\exp\left\{[\cL_0 + \cG_{\mathrm{Z}}(\tau)]t\right\} .
\label{ZenoApp}
\end{eqnarray}
Under the assumption that the bath state can well be approximated
by an equilibrium state at time $t$, the final reduced state
$\sigma(t)$ is shown to satisfy the equation
\begin{eqnarray}
\dot \sigma(t) =  \left[\cL_S +\cL_{\mathrm Z}(\tau)\right]\sigma(t) \ ,
\label{sig*Eq}
\end{eqnarray}
with
\begin{eqnarray}
\cL_{\mathrm Z}(\tau) \sigma =
\mathrm{tr}_B
\left\{\cG_{\mathrm{Z}}(\tau) \sigma\otimes \rho_B\right\} \ .
\label{cLZ}
\end{eqnarray}
Note that $\cL_{\mathrm{Z}}(\tau)$ is the solution of the operator
equation
\begin{equation}\label{eq:cLZ}
\int_0^\tau dt\; e^{-\cL_S t}\cL_{\mathrm{Z}}(\tau)e^{\cL_S t}=
\int_0^\tau dt {\hat P} \cL(t){\hat P}= \int_0^\tau dt\int_0^t
ds\; {\hat P}\cK_{\mathrm{I}}(t,s){\hat P} \ .
\end{equation}
where $\cK_{\mathrm{I}}$ is defined in (\ref{eq:cKI}). The
dissipative part of (\ref{cLZ}) is found to have the explicit form
[analogous to Eq.\ (\ref{eq:Liouvilliannrw})]
\begin{eqnarray}
\cL_{\mathrm{Z}}(\tau) \sigma &=& \gamma_0^{\mathrm{Z}}(\tau){\hat
P}\left(X_0{\hat P}\sigma X_0-\frac{1}{2}\left\{X_0
X_0, {\hat P}\sigma\right\}\right)\nonumber\\
& & + \sum_{m\ge1} \gamma_m^{\mathrm{Z}}(\tau){\hat
P}\left(X_m{\hat P}\sigma
X_m^\dagger-\frac{1}{2}\left\{X_m^\dagger X_m, {\hat
P}\sigma\right\}\right)+ \sum_{m\ge1}
\gamma_{-m}^{\mathrm{Z}}(\tau){\hat P} \left(X_m^\dagger{\hat
P}\sigma X_m-\frac{1}{2}\left\{X_m X_m^\dagger, {\hat
P}\sigma\right\}\right) \ , \nonumber \\
\label{eq:disscLz}
\end{eqnarray}
where the controlled decay rates read
\begin{equation}\label{eq:gammazenodef}
\gamma_{m}^{\mathrm{Z}}(\tau)= \frac{2}{\tau} \Re \int_0^\tau dt
\int_{-t}^0 ds \left(\left\langle A_m A^\dagger_m(s)\right\rangle+
\left\langle A_{-m}^\dagger A_{-m}(s)\right\rangle\right) =\tau
\int_{-\infty}^\infty d \omega \; \kappa_m^\beta(\omega)\;
 {\rm sinc}^2 \left(\frac{\omega-\omega_m}{2}
\tau\right)
 \ ,
\end{equation}
with $\mathrm{sinc}(x)=\sin(x)/x$. This yields Zeno and inverse
Zeno effects as $\tau$ is changed, as we will see in Sec.\
\ref{sec:examples}. The key issue, once again, is to understand
\emph{how small} $\tau$ should be in order to get suppression
(control) of decoherence (QZE), rather than its enhancement (IZE).

\section{Control via dynamical decoupling}
\label{sec.kickb} We can now investigate the nonideal bang-bang
control of decoherence. From Eq.\ (\ref{eq:limcLk}), describing a
BB control with a single kick \cite{bang},
\begin{equation}
\label{noname0} \rho(t) = \left[e^{\cL_\mathrm{k}} e^{{\cal
L}_{\rm tot} \tau} \right]^{\frac{t}{\tau}}   \rho(0) \ ,
\end{equation}
where $\cL_{\mathrm{tot}}$ is again given by (\ref{Ltotdef}). As
in the Zeno control, we consider here the case where $\tau$ is
finite, so that the effects on the decay rate arising from the
presence of a short-time quadratic (Zeno) region play a
fundamental role. Once again, we only sketch the main steps in the
derivation and give more details in Appendix \ref{sec:C}. Second
order perturbation in $\cL_{SB}$ yields
\begin{eqnarray}
e^{\cL_\mathrm{k}} e^{{\cal L}_{\rm tot} \tau}  =
e^{\cL_\mathrm{k}} e^{\cL_0 \tau} {\cal T} \exp\left(\int_0^{\tau}
ds\; \cL_{SB}(s)\right)  \simeq e^{\cL_\mathrm{k}} e^{\cL_0 \tau}
\left[ \openone + \int_0^\tau ds\; \cL_{SB}(s) + \int_0^\tau
ds\int_0^s ds_1\; \cL_{SB}(s)\cL_{SB}(s_1) \right] \ .
\end{eqnarray}
In terms of the operators $\cF_\mathrm{k}(\tau)$ and
$\cG_{\mathrm{k}}(\tau)$, defined as solutions of the operator
equations
\begin{eqnarray}
\label{noname1}
& & \int_0^{\tau} ds\; e^{-\cL_\tau s} \cF_\mathrm{k} (\tau)
e^{\cL_\tau s}= \int_0^\tau ds\; \cL_{SB}(s)
\\
& & \int_0^{\tau} ds\; e^{-\cL_\tau s} \cG_\mathrm{k} (\tau)
e^{\cL_\tau s}= \int_0^\tau ds\int_0^s ds_1\;
\left[\cL_{SB}(s)\cL_{SB}(s_1)- e^{-\cL_\tau s}\cF_\mathrm{k}
(\tau) e^{\cL_\tau (s-s_1)}\cF_\mathrm{k} (\tau) e^{\cL_\tau
s_1}\right]
 \ ,
 \label{noname2}
\end{eqnarray}
with
\begin{equation}
\cL_\tau=\frac{\cL_\mathrm{k}}{\tau}+\cL_0,
\end{equation}
one has
\begin{eqnarray}
\left[e^{\cL_\mathrm{k}} e^{{\cal L}_{\rm tot} \tau} \right]^N
\simeq \left[e^{\cL_\tau \tau}  {\cal T} \exp\left(\int_0^{\tau}
ds\; e^{-\cL_\tau s}
\left(\cF_\mathrm{k}(\tau)+\cG_\mathrm{k}(\tau)\right) e^{\cL_\tau
s}\right)\right]^N=
\exp\left\{\left[\frac{\cL_\mathrm{k}}{\tau}+\cL_0+
\cF_\mathrm{k}(\tau)+\cG_\mathrm{k}(\tau)\right]t\right\} .
\label{kickApp}
\end{eqnarray}
With the aid of (\ref{kickApp}), the final reduced state
$\sigma(t)$ satisfies the equation
\begin{eqnarray}
\dot \sigma(t) =  \left[\frac{\cL_{k}}{\tau}+\cL_S
+\cL_{\mathrm{k}}(\tau)\right]\sigma(t) \ ,
\label{sig*Eq1}
\end{eqnarray}
with
\begin{eqnarray}
\cL_{\mathrm{k}}(\tau) \sigma = \mathrm{tr}_B
\left\{\cG_\mathrm{k}(\tau) \sigma \otimes \rho_B\right\}\ .
\label{cLk}
\end{eqnarray}
The dissipative part of (\ref{cLk}) has the explicit form
\begin{eqnarray}
\cL_{\mathrm{k}}(\tau) \sigma &=&
\gamma_0^{\mathrm{k}}(\tau)\left(X_0(\tau)\sigma
X_0(\tau)-\frac{1}{2}\left\{X_0(\tau) X_0(\tau),
\sigma\right\}\right)+ \sum_{m\ge1}
\gamma_m^{\mathrm{k}}(\tau)\left(X_m(\tau)\sigma X_m^\dagger
(\tau)-\frac{1}{2}\left\{X_m^\dagger(\tau) X_m(\tau),
\sigma\right\}\right)\nonumber\\
& & + \sum_{m\ge1} \gamma_{-m}^{\mathrm{k}}(\tau)
\left(X_m^\dagger(\tau)\sigma
X_m(\tau)-\frac{1}{2}\left\{X_m(\tau) X_m^\dagger(\tau)
\sigma\right\}\right) \ ,
\label{eq:disscLk}
\end{eqnarray}
where, in analogy with Eq.\ (\ref{eq:eigenoperators}), the
$X_m(\tau)$ are the eigenoperators of the Liouvillian
$\cL_\mathrm{k}/\tau+\cL_S$, satisfying
\begin{equation}\label{eq:eigenoperatorstau}
\left(\frac{\cL_\mathrm{k}}{\tau}+\cL_S\right) X_m(\tau)= i
\omega_m(\tau) X_m(\tau) \qquad (\omega_m\neq\omega_n, \quad
\mathrm{for}\quad m\neq n)
\end{equation}
and the controlled decay rates read
\begin{eqnarray}
\gamma_{m}^{\mathrm{k}}(\tau)= 2 \Re \int_0^\infty dt
\left(\left\langle \tilde A_m(0) \tilde
A^\dagger_m(t)\right\rangle+ \left\langle \tilde A_{-m}^\dagger(0)
\tilde A_{-m}(t)\right\rangle\right) = 2\pi
\kappa_m^\beta\left(\omega_m(\tau)\right)= 2\pi
\kappa_m^\beta\left(\frac{2\pi m}{\tau} +\mathrm{O}(1)\right)\ ,
\label{eq:gammakick}
\end{eqnarray}
with
\begin{equation}\label{eq:Amtilde}
\tilde A_m(t)=\int d^3 k g_m^*({\bm k})
e^{-i(\omega_k-\omega_m(\tau))t} a({\bm k}) .
\end{equation}

Notice that the mechanism of decoherence suppression
(\ref{eq:gammakick}) is not fully determined by $\cL_{\rm tot}$
and $\hat P$, in contrast to the Zeno case, and depends also on
the details of the Liovillian $\cL_{\mathrm{k}}$ through $\omega_m
(\tau)$. This is best clarified by explicitly looking at a
particular case: let us consider the two level system
(\ref{eq:2Hamiltonian}) with $g_0=0$ (spin-flip decoherence). We
include an additional third level---that performs the
control---and add to (\ref{eq:2Hamiltonian}) the Hamiltonian
(acting on $\mathcal{H}_{S}\oplus \mathrm{span}\{|M\rangle\}$)
\begin{equation}\label{eq:thirdlev}
H_M = -\frac{\Omega}{2} |M\rangle \langle M|,
\end{equation}
so that $|M\rangle$ is degenerate with $|{\downarrow}\rangle$. The
control consists of a sequence of $2\pi$ pulses
\cite{ShiokawaLidar} between $|{\downarrow}\rangle$ and
$|M\rangle$, given by
\andy{hmeask}
\barr
U_{\mathrm{k}} = \exp\left[-i \pi \left( |{\downarrow}\rangle
\langle M| + |M\rangle \langle{\downarrow}|\right)\right] =
P_{\uparrow}-P_{-1} \; ,
\label{eq:hmeask}
\earr
where
\begin{equation}
P_{\uparrow}=|\uparrow\rangle\langle\uparrow|,\qquad
P_{-1}=P_{\downarrow}+P_M=|\downarrow\rangle\langle\downarrow|+
|M\rangle\langle M|,
\label{eq:meassubbk}
\end{equation}
are the eigenprojections of $U_{\mathrm{k}}$ (belonging
respectively to $e^{-i\lambda_\uparrow}=1$ and
$e^{-i\lambda_{-1}}=-1$) which define two Zeno subspaces. In the
$\tau\to0$ limit any decoherence between these two subspaces is
suppressed. In fact, the total decay rate of the upper level has
been explicitly computed \cite{BBDDsem,ShiokawaLidar} and reads
\begin{equation}\label{eq:gammakickViola}
\gamma_{\uparrow}^{\mathrm{k}}(\tau)= \lim_{t\to\infty} t
\int_{-\infty}^\infty d \omega \; \kappa^\beta(\omega)\;
 {\rm sinc}^2 \left(\frac{\omega-\Omega}{2}\,
t\right) \tan^2 \left(\frac{\omega-\Omega}{2}\, \tau \right)
 \ .
\end{equation}
As a matter of fact, the function multiplying the thermal form
factor inside the integral can be shown to have the interesting
limit
\begin{equation}\label{eq:intkick}
\lim_{t\to\infty} t\; {\rm sinc}^2 \left(\frac{\omega t}{2}\right)
\tan^2 \left(\frac{\omega \tau}{2} \right)=
\frac{2}{\pi}\sum_{j=0}^\infty
\frac{1}{\left(j+\frac{1}{2}\right)^2}
 \left[\delta\left(\omega-\frac{2\pi}{\tau}(j+1/2)\right)
+\delta\left(\omega+\frac{2\pi}{\tau}(j+1/2)\right)\right] \ .
\end{equation}
The above limit is taken by keeping $\tau$ fixed---finite and
nonvanishing---and $t=N\tau$, with $N$ integer and even
\cite{ShiokawaLidar}. By plugging (\ref{eq:intkick}) into
(\ref{eq:gammakickViola}) one gets
\begin{equation}\label{eq:gammakickdef}
\gamma^{\mathrm{k}}_{\uparrow}(\tau)=
\frac{2}{\pi}\sum_{j=0}^\infty
\frac{1}{\left(j+\frac{1}{2}\right)^2}
 \left[\kappa^\beta\left(\Omega+\frac{2\pi}{\tau}(j+1/2)\right)
+\kappa^\beta\left(\Omega-\frac{2\pi}{\tau}(j+1/2)\right)\right] \
,
\end{equation}
which is a sum of suitably weighted terms of the form
(\ref{eq:gammakick}). This yields again control of decoherence as
$\tau$ is varied, as we will see in Sec.\ \ref{sec:examples}. The
key issue, once again, is to understand \emph{how small} $\tau$
should be in order to get suppression of decoherence (control),
rather than its enhancement. Equation (\ref{eq:gammakickdef})
yields also a significant computational advantage, when compared
to (\ref{eq:gammakickViola}): for well-behaved form factors
(without resonances) the first few terms already provide a good
estimate of the controlled lifetime.

\section{Control via a strong continuous coupling}
\label{sec:dyndec}

We can now analyze the last case, that of control by means of a
strong continuous coupling. Since the control of decoherence is
achieved by adding a control Hamiltonian $KH_{\mathrm{c}}$ acting
on the Hilbert space $\mathcal{H}_{S}$, we begin with the study of
the spectral properties of the new ``system'' Hamiltonian $H_S(K)
\equiv H_S+KH_{\mathrm{c}}$. By writing the spectral resolutions
of $H_S$ and $H_\mathrm{c}$,
\begin{equation}\label{eq:specHSHc}
H_S = \sum_n E_n Q_n, \qquad H_\mathrm{c} = \sum_m \eta_m P_m,
\end{equation}
with $\sum_n Q_n=\sum_m P_m=\openone$, and by using the property
(\ref{eq:nozeno1}) we see that $P_{mn}=P_m Q_n$ is a (finer)
orthogonal resolution of the identity, i.e. $\sum_{m,n}
P_{mn}=\openone$, with
$P_{mn}P_{m'n'}=\delta_{m,m'}\delta_{n,n'}P_{mn}$. Note that some
$P_{mn}$ can vanish. In particular $H'_S$ can be explicitly
diagonalized
\begin{equation}\label{eq:diagH1S}
H_S(K)= \sum_{m,n} \left( K \eta_m + E_n\right) P_{mn}, \qquad
P_{mn}=P_m Q_n \ .
\end{equation}
Equations (\ref{eq:specHSHc}) and (\ref{eq:diagH1S}) directly
translate in terms of Liouvillian as
\begin{equation}\label{eq:specLSLc}
\cL_S = -i\sum_n \omega_n {\tilde Q}_n, \qquad \cL_\mathrm{c} =
-i\sum_m \Omega_m {\tilde P}_m
\end{equation}
and
\begin{eqnarray}
\cL_S(K)&=&\cL_S +
K\cL_\mathrm{c}=-i\sum_{m,n}\omega_{mn}(K){\tilde P}_{mn},
\nonumber\\
\omega_{mn}(K) &=& K \Omega_m + \omega_n,
 \qquad {\tilde P}_{mn}={\tilde P}_m{\tilde Q}_n
\ .
\label{eq:diagL1S}
\end{eqnarray}
The condition (\ref{eq:C}) for a complete control of decoherence,
${\hat P}H_{SB}=0$, leads to
\begin{equation}
0={\hat P} H_{SB}=\sum_m P_m H_{SB} P_m=\tilde P_0 H_{SB}=\sum_n
\tilde P_0 \tilde Q_n H_{SB}= \sum_n \tilde P_{0n} H_{SB} ,
\label{eq:1/KExp3}
\end{equation}
whence
\begin{equation}\label{eq:tildeP0n}
\tilde P_{0n} H_{SB}=0, \qquad \forall n \ .
\end{equation}
Therefore, by following exactly the same steps of
Sec.~\ref{sec.modelA}, with $H_S(K)$ defined by (\ref{eq:diagH1S})
in place of $H_S$, one obtains that the dissipative part of the
Liouvillian $\cL_K$ governing the slow evolution of the reduced
density matrix $\sigma$ is given by
\begin{eqnarray}
\cL_K \sigma &=&  \sum_{m\ge1,\, n}
\gamma_{mn}(K)\left(X_{mn}\sigma
X_{mn}^\dagger-\frac{1}{2}\left\{X_{mn}^\dagger X_{mn},
\sigma\right\}\right)\nonumber\\
& & + \sum_{m\ge1, \,n} \gamma_{-m,n}(K)
\left(X_{mn}^\dagger\sigma X_{mn}-\frac{1}{2}\left\{X_{mn}
X_{mn}^\dagger, \sigma\right\}\right) \ ,
\label{eq:LiouvillianStrong}
\end{eqnarray}
where
\begin{equation}\label{eq:XmnK}
X_{mn}\equiv\tilde P_{m}X_n \ ,
\end{equation}
with $X_n$ given by (\ref{eq:intHam})-(\ref{eq:eigenoperators}),
and
\begin{equation}\label{eq:gammastrongdef}
\gamma_{mn}(K)= 2\Re \int_0^\infty dt \left(\left\langle \tilde
A_{mn}(0)\tilde A^\dagger_{mn}(t)\right\rangle+ \left\langle
\tilde A_{-m,n}^\dagger(0)\tilde A_{-m,n}(t)\right\rangle\right) =
2\pi \kappa_n^\beta\left(\omega_{mn}(K)\right)
 \ ,
\end{equation}
with
\begin{equation}\label{eq:AtildeK}
 \tilde A_{mn}(t)=\int d^3k\;  g_n^*({\bm k})
e^{-i(\omega_k-\omega_{mn}(K))t} a({\bm k}) \ .
\end{equation}
All terms with $m=0$ identically vanish due to
(\ref{eq:tildeP0n}). In the $K\to\infty$ limit, because the
thermal form factor $\kappa_m^\beta(\omega)$ vanishes as
$\omega\to \infty$ (cf.\ Fig.\ 2), one has
\begin{equation}\label{eq:1/KExp2}
\gamma_{mn}(K) = 2\pi\kappa_n^\beta\left(K \Omega_m +
\omega_n\right) \sim 2\pi\kappa_n^\beta\left(K \Omega_m \right)
\to 0 \ , \qquad \mbox{for} \quad K\to\infty \ .
\end{equation}
Hence, in the $K\to +\infty$ limit, the dissipative part
disappears, $\cL_K \to 0$, or decoherence is suppressed, as
expected.

It is interesting to observe that, when the condition
(\ref{eq:nozeno1}) is not satisfied, the control via a strong
continuous coupling needs an additional argument. In such a case,
the control Hamiltonian $H_c$ and the system Hamiltonian $H_S$
cannot be simultaneously diagonalized, but (for a
finite-dimensional ${\cal H}_S$), as a result of the analyticity
of the eigenvalues and the corresponding eigenprojections of the
Hermitian operator $H_S(K)/K=H_S/K+H_c$ with respect to the
perturbation parameter $1/K$ \cite{KatoLinearOperators}, the
eigenvalues $\omega_{mn}(K)$ of the new system Liouvillian ${\cal
L}_S(K)=K{\cal L}_c+{\cal L}_S$ and the corresponding
eigenprojections ${\tilde P}_{mn}(K)$ satisfy
\begin{eqnarray}
\omega_{mn}(K)&=& K\Omega_m + \Omega_{mn}^{(1)} + {\rm O}
\left({1\over K}\right) \ ,\label{eq:1/KExp1} \\
{\tilde P}_{mn}(K)&=& {\tilde P}_{mn}^{(0)} + {1\over K}{\tilde
P}_{mn}^{(1)} + {\rm O} \left({1\over K^2}\right)\ ,
\end{eqnarray}
where $\Omega_{mn}^{(1)}$ and ${\tilde P}_{mn}^{(j)}$ ($j=0,1$) do
not depend on $K$. As in (\ref{eq:tildeP0n}), one gets that
${\tilde P}_{0n}^{(0)}H_{SB}=0$, but this does not imply that
${\tilde P}_{0n}(K)H_{SB}=0$. As a result, there appear
dissipative terms which tend to 0 via a different mechanism from
the one outlined above. This aspect will be discussed elsewhere,
together with similar phenomena that occur also for the other two
control mechanisms (BB and Zeno).

In general, as in the BB control but in contrast to the Zeno case,
the mechanism of decoherence suppression (\ref{eq:1/KExp2}) is not
fully determined by $H_S$ and depends on the details of the
Hamiltonians $H_S$ \emph{and} $H_{\mathrm{c}}$. Once again, this
can be clarified by looking at a specific example: consider the
two level system (\ref{eq:2Hamiltonian}) with $g_0=0$ (spin flip
decoherence). We add to (\ref{eq:2Hamiltonian}) the Hamiltonian
(acting on $\mathcal{H}_{S}\oplus \mathrm{span}\{|M\rangle\}$)
\andy{hmeas}
\barr H_M &=& -\frac{\Omega}{2} |M\rangle \langle M|+
KH_{\mathrm{c}},
\nonumber \\
H_{\mathrm{c}} &=&  |{\downarrow}\rangle \langle M| + |M\rangle
\langle{\downarrow}| =P_{+}-P_{-} \; ,
\label{eq:hmeas}
\earr
where
\begin{equation}
P_\pm=\frac{(|{\downarrow}\rangle \pm |M\rangle)(\langle{\downarrow}|\pm
\langle M|)}{2}\equiv|\pm\rangle\langle\pm|.
\label{eq:meassubb}
\end{equation}
The third state $|M\rangle$ is now ``continuously" coupled to
state $|{\downarrow}\rangle$, $K \in \mathbb{R}$ being the
strength of the coupling. As $K$ is increased, state $|M\rangle$
performs a better ``continuous observation" of
$|{\downarrow}\rangle$, yielding the Zeno subspaces \cite{PIO}. In
terms of its eigenprojections, $H_{\mathrm{c}}$ reads [see
(\ref{eq:diagevol})] \beq H_{\mathrm{c}}= \eta_{\uparrow}
P_{\uparrow}+ \eta_{-} P_{-} + \eta_{+} P_{+} , \eeq with
$P_{\uparrow}=|{\uparrow}\rangle\langle{\uparrow}|$ and
$\eta_\uparrow=0, \eta_\pm = \pm 1$. In the Zeno limit ($K\to
\infty$) the subspaces $\mathcal{H}_{\uparrow}$, $\mathcal{H}_{+}$
and $\mathcal{H}_{-}$ decouple due to wildly oscillating phases
O$(K)$. We get
\begin{equation}
\hat{P} H_{SB} = P_{\uparrow} H_{SB} P_{\uparrow}+ P_{-} H_{SB}
P_{-} + P_{+} H_{SB} P_{+}=0 .
\end{equation}
Therefore in the limit $K\to\infty$, $\gamma_{\pm1}=0$ and
decoherence is halted.

We can diagonalize the new system Hamiltonian
\begin{equation}\label{eq:diagonal}
H_S'=\frac{\Omega}{2} \sigma_z -\frac{\Omega}{2} |M\rangle
\langle M|+ KH_{\mathrm{c}}= \frac{\Omega}{2} P_{\uparrow}+
\left(-\frac{\Omega}{2}+K\right)P_{+} +
\left(-\frac{\Omega}{2}-K\right)P_{-} \ .
\end{equation}
The new system operators (\ref{eq:X+-}) become
\begin{equation}\label{eq:X+-dressed}
X_{\pm}=P_{\pm}\sigma_x
P_{\uparrow}=\frac{1}{\sqrt{2}}|\pm\rangle\langle{\uparrow}|,
\quad X_0=|-\rangle\langle+|,\qquad
\cL_{S}' X_{\pm}=i(\Omega\mp K) X_{\pm} \ ,\quad
\cL_{S}' X_0=2iKX_0,
\end{equation}

\begin{equation}\label{eq:dressHSB}
H_{SB}=\left(X_{+}+X_{-}+X_{+}^\dagger+X_{-}^\dagger\right)\otimes\left(
A(g)+A^\dagger(g)\right) ,
\end{equation}
hence
\begin{eqnarray}
\cL_K \rho &=& \gamma_{+}(K)\left(X_{+} \rho X_{+}^\dagger
-\frac{1}{2}\left\{X_{+}^\dagger X_{+}, \rho\right\}\right)+
\gamma_{-}(K)\left(X_{-} \rho X_{-}^\dagger
-\frac{1}{2}\left\{X_{-}^\dagger X_{-}, \rho\right\}\right)
\nonumber\\
& + &  \bar\gamma_{+}(K)\left(X_{+}^\dagger \rho X_{+}
-\frac{1}{2}\left\{X_{+} X_{+}^\dagger, \rho\right\}\right)+
\bar\gamma_{-}(K)\left(X_{-}^\dagger \rho X_{-}
-\frac{1}{2}\left\{X_{-} X_{-}^\dagger, \rho\right\}\right) \ ,
\label{eq:Liouvillian2ldress}
\end{eqnarray}
where
\begin{equation}\label{eq:gammapm}
 \gamma_{\pm}(K)= 2\pi \kappa_1^\beta (\Omega\mp K), \qquad
 \bar\gamma_{\pm}(K)= 2\pi \kappa_1^\beta (-\Omega\pm K)\ .
\end{equation}
For example, the decay rate out of state $|{\uparrow}\rangle$
reads
\begin{equation}\label{eq:dressup}
 \gamma_{\uparrow}(K)= \frac{\gamma_{+}(K)+\gamma_{-}(K)}{2}=
 \pi \left(\kappa_1^\beta (\Omega - K)+\kappa_1^\beta (\Omega +
 K)\right).
\end{equation}

\section{The role of the form factors}
\label{sec:examples}

We can now test the general scheme described in the previous
sections by looking in detail at some particular cases. We will
consider the two-level situation and compare the three control
methods both with exponential (\ref{eq:expformfactor}) and
polynomial form factors (\ref{eq:lorformfactor1}). We will
concentrate on the transition between a regime in which
decoherence is partially suppressed (``controlled") and a regime
in which it is enhanced. We shall work in a high-temperature
regime, which is rather critical from an experimental point of
view, because of temperature-induced transitions in two-level
systems. We shall set $\Omega=0.01W$ and $\beta=50W^{-1}$, so that
temperature=$\beta^{-1}=2\Omega$.

\subsection{Quantum Zeno control}

\begin{figure}
\begin{center}
\includegraphics[width=\figwidth]{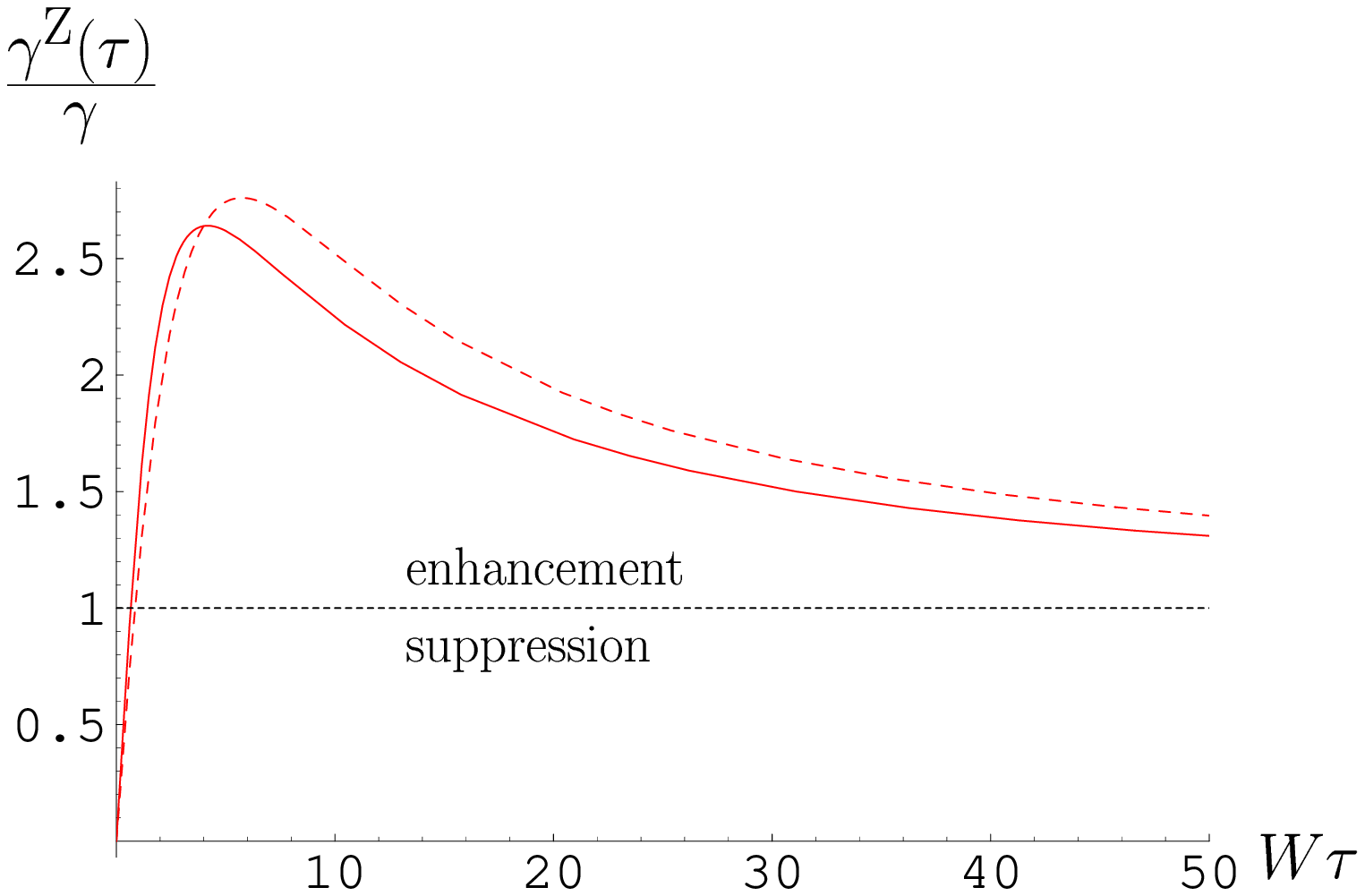}
\end{center}
\caption{Projective measurements:
$\gamma^{\mathrm{Z}}(\tau)/\gamma$ vs $W\tau$. Full line:
exponential form factor (\ref{eq:expformfactor}); dashed line:
polynomial form factor (\ref{eq:lorformfactor1}) with $n=2$.}
\label{fig:meas}
\end{figure}

\begin{figure}
\begin{center}
\includegraphics[width=\figwidth]{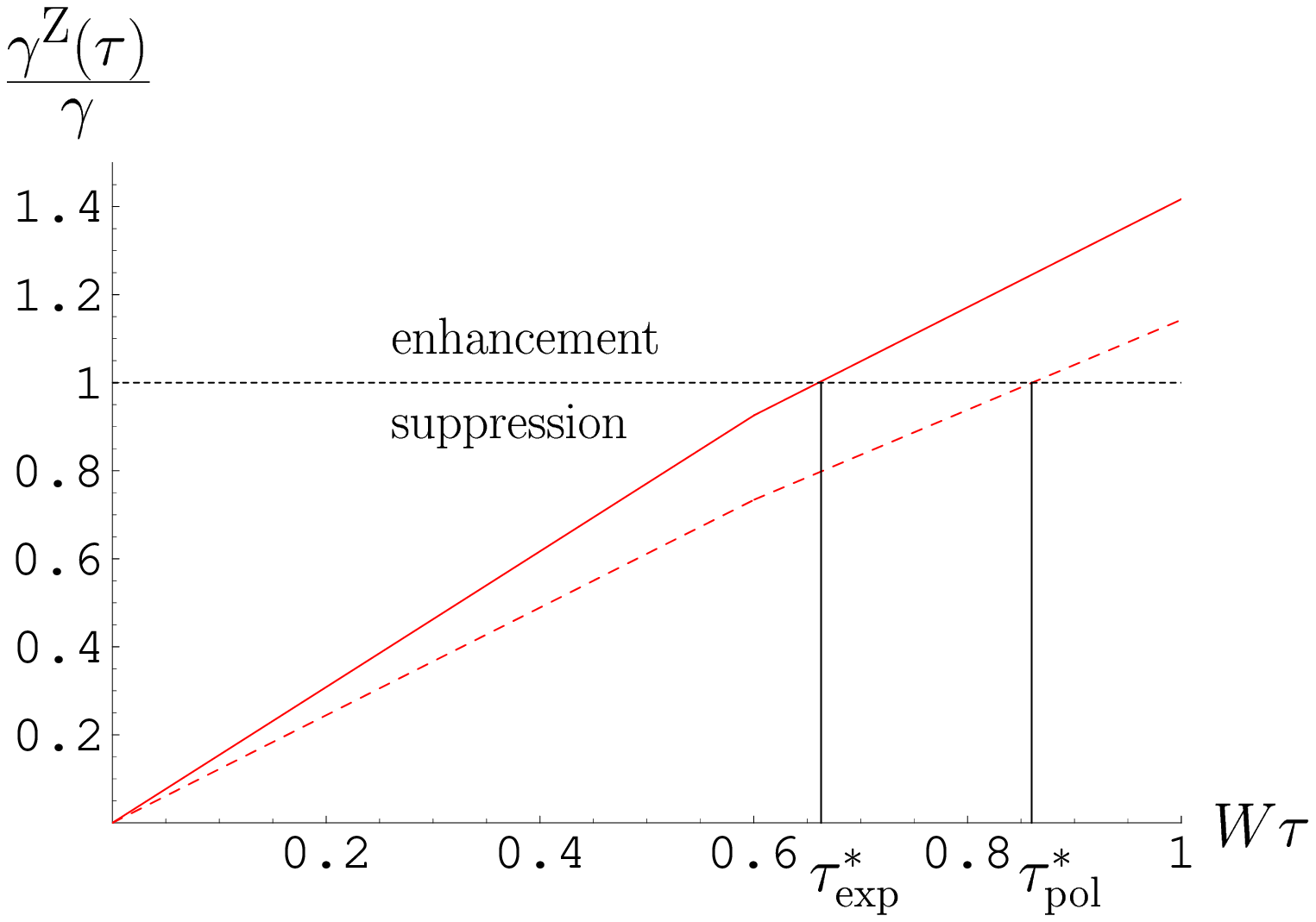}
\end{center}
\caption{Projective measurements:
$\gamma^{\mathrm{Z}}(\tau)/\gamma$ vs $W\tau$, for small $\tau$.
Full line: exponential form factor (\ref{eq:expformfactor});
dashed line: polynomial form factor (\ref{eq:lorformfactor1}) with
$n=2$. $\tau^*$ (indicated) is defined by the equation
$\gamma^{\mathrm{Z}}(\tau^*)/\gamma=1$. Decoherence is suppressed
when $\gamma^{\mathrm{Z}}(\tau)<\gamma$; it is enhanced
otherwise.}
\label{fig:meassmall}
\end{figure}

We first consider the Zeno control by projective measurements.
Dissipation and decoherence are characterized by the decay rate
(\ref{eq:gammazenodef}):
\begin{equation}\label{eq:gammazenolim}
\gamma^{\mathrm{Z}}(\tau)= \tau \int_{-\infty}^\infty d \omega \;
\kappa^\beta(\omega)\;
 {\rm sinc}^2 \left(\frac{\omega-\Omega}{2}
\tau\right) \sim \frac{\tau}{\tau_{\mathrm{Z}}^2}, \qquad \tau\to
0 ,
\end{equation}
where $\tau_{\mathrm{Z}}$,
\begin{equation}\label{eq:tauZtherm}
\tau_{\mathrm{Z}}^{-2}=\int_{-\infty}^\infty d \omega \;
\kappa^\beta(\omega)=\int_0^\infty d\omega\; \kappa(\omega) \coth
\left(\frac{\beta\omega}{2}\right)\stackrel{\beta\to\infty}{\thicksim}
\int_0^\infty d\omega\; \kappa(\omega)+ 2\int_0^\infty d\omega\;
\kappa(\omega) \exp(-\beta\omega),
\end{equation}
is the thermal Zeno time. (We dropped the suffix $m$ for
simplicity.) Observe that, by making use of the limit
\begin{equation}\label{eq:sinclim}
\lim_{\tau\to\infty} \tau\; \mathrm{sinc}^2
\left(\frac{\omega\tau}{2}\right)=2\pi \delta(\omega) \,
\end{equation}
one gets
\begin{equation}\label{eq:gammazenolim1}
\gamma^{\mathrm{Z}}(\tau) \to \gamma, \qquad \tau\to \infty
 \ ,
\end{equation}
where
\begin{equation}
\gamma= 2\pi \kappa^\beta(\Omega)
\label{eq:freegamma}
\end{equation}
is the natural decay rate (\ref{eq:gammamdef}). The ratio
$\gamma^{\mathrm{Z}}(\tau)/\gamma$ is the key quantity:
decoherence is suppressed if $\gamma^{\mathrm{Z}}(\tau)<\gamma$,
and it is enhanced otherwise. This ratio is shown in Fig.\
\ref{fig:meas} as a function of $\tau$ [in units $W$--the bandwidth
defined in Eq.\ (\ref{eq:BW})].
The transition between these two regimes takes place at
$\tau=\tau^*$, where $\tau^*$ is defined by the equation
\cite{Heraclitus}
\begin{equation}\label{eq:tau*def}
 \gamma^{\mathrm{Z}}(\tau^*) = \gamma^{\mathrm{Z}}(\infty)=\gamma .
\end{equation}
If $\tau^*$ belongs to the linear region (\ref{eq:gammazenolim})
(which is our case and is true for sufficiently small energy
$\Omega$ of the initial state), one gets
\begin{equation}\label{eq:tjump}
\tau^* \simeq \gamma \tau_{\mathrm{Z}}^2= 2\pi
\frac{\kappa^\beta(\Omega)}{\int_{-\infty}^\infty d \omega \;
\kappa^\beta(\omega)}.
\end{equation}
The short time region is displayed for clarity in Fig.\
\ref{fig:meassmall}.

It is useful to spend a few words on the \emph{physical} meaning
of the expressions $\tau \to 0$, $\beta \to \infty$ in the above
(and following) formulas. Times and temperatures are to be
compared with the bandwidth $W$ (or frequency cutoff $\Lambda$).
Times (temperatures) are ``small" when $\tau \ll W^{-1}$
($\beta^{-1}
\ll W$). For example, it is worth emphasizing that
the relevant timescale is $\tau^*$, when one considers short-time
expansions in a Zeno context \cite{Heraclitus,Antoniou}: the
expansion (\ref{eq:gammazenolim}) is valid for $\tau \lesssim
W^{-1}$ (and not $\tau \lesssim \tau_{\rm Z}$, as it is sometimes
erroneously assumed).

\subsection{``Bang bang" Control}

We now discuss BB. The decay rate is given by Eq.\
(\ref{eq:gammakickdef}):
\begin{eqnarray}
 \gamma^{\mathrm{k}}(\tau) &=& \frac{2}{\pi}\sum_{j=0}^\infty
\frac{1}{\left(j+\frac{1}{2}\right)^2}
 \left[\kappa^\beta\left(\Omega+\frac{\pi}{\tau}(2j+1)\right)
+\kappa^\beta\left(\Omega-\frac{\pi}{\tau}(2j+1)\right)\right]
\nonumber\\ &\stackrel{\tau\to 0}{\thicksim}&
\frac{2}{\pi}\sum_{j=0}^\infty
\frac{1}{\left(j+\frac{1}{2}\right)^2}\;
\kappa^\beta\left(\frac{\pi}{\tau}(2j+1)\right)
 \left(1+e^{-\beta \frac{\pi}{\tau}(2j+1)}\right)
\thicksim \frac{2}{\pi}\sum_{j=0}^\infty
\frac{1}{\left(j+\frac{1}{2}\right)^2}\;
\kappa\left(\frac{\pi}{\tau}(2j+1)\right) ,
 \label{eq:gammaklim}
\end{eqnarray}
where we made use of (\ref{eq:excsym}) in the first expansion and
assumed that $\beta$ is not too small (as compared to $\tau$) in
the second one. In the exponential case (\ref{eq:expformfactor})
one gets
\begin{equation}
\kappa^{(E)}\left(\frac{\pi}{\tau}(2j+1)\right) =
g^2\frac{\pi}{\tau}(2j+1)e^{-\frac{\pi}{\tau \Lambda}(2j+1)} =
\kappa^{(E)}\left(\frac{\pi}{\tau}\right)(2j+1) e^{-2j
\frac{\pi}{\tau\Lambda}}\ ,
\end{equation}
whence
\begin{eqnarray}
 \gamma^{\mathrm{k}}(\tau)
 \sim \frac{8}{\pi}\,
\kappa^{(E)}\left(\frac{\pi}{\tau}\right) , \qquad \tau \to 0 \ ,
 \label{eq:gammaklimexp}
\end{eqnarray}
while in the polynomial case (\ref{eq:lorformfactor1}) one gets
\begin{equation}
\kappa^{(P)}\left(\frac{\pi}{\tau}(2j+1)\right) \sim
g^2\frac{\Lambda}{\left[\frac{\pi}{\tau\Lambda}(2j+1)\right]^{2n-1}}
\sim
\kappa^{(P)}\left(\frac{\pi}{\tau}\right)\frac{1}{(2j+1)^{2n-1}} \
,
\end{equation}
whence
\begin{eqnarray}
 \gamma^{\mathrm{k}}(\tau)
 \sim \frac{8}{\pi} \sum_{j=0}^\infty
\frac{1}{(2j+1)^{2n+1}} \,
\kappa^{(P)}\left(\frac{\pi}{\tau}\right)=
\frac{8}{\pi}\left(1-2^{-2n-1}\right)\zeta(2n+1)\,
\kappa^{(P)}\left(\frac{\pi}{\tau}\right) , \qquad \tau \to 0 ,
 \label{eq:gammaklimpow}
\end{eqnarray}
where $\zeta(x)=\sum_{k=1}^{\infty} k^{-x}$ is the Riemann zeta
function.

\begin{figure}
\begin{center}
\includegraphics[width=\figwidth]{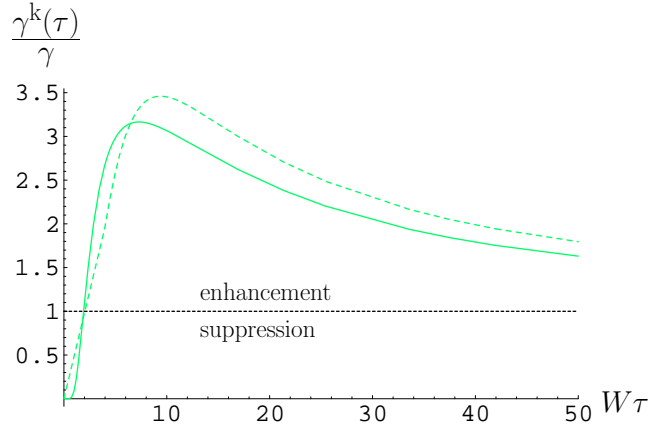}
\end{center}
\caption{BB kicks:
$\gamma^{\mathrm{k}}(\tau)/\gamma$ vs $W\tau$. Full line:
exponential form factor (\ref{eq:expformfactor}); dashed line:
polynomial form factor (\ref{eq:lorformfactor1}) with $n=2$.}
\label{fig:kick}
\end{figure}

On the other hand, in both cases,
\begin{eqnarray}
 \gamma^{\mathrm{k}}(\tau) \to
 \frac{4}{\pi}\,\kappa^\beta\left(\Omega\right) \sum_{j=0}^\infty
\frac{1}{\left(j+\frac{1}{2}\right)^2}= \gamma, \qquad
\tau\to\infty ,
\label{eq:gammaklim1}
\end{eqnarray}
where we summed the series
\begin{equation}\label{eq:sumser}
 \sum_{j=0}^\infty \frac{1}{\left(j+\frac{1}{2}\right)^2}
 = 4 \sum_{j=0}^\infty \frac{1}{\left(2j+1\right)^2}= 3\, \zeta(2)=
 \frac{\pi^2}{2} \ .
\end{equation}
The ratio $\gamma^{\mathrm{k}}(\tau)/\gamma$ is shown in Fig.\
\ref{fig:kick} as a function of $\tau$.
Once again, the transition between the two regimes takes place at
$\tau=\tau^*$ where $\tau^*$ is defined by the equation
\begin{equation}\label{eq:tau*def1}
 \gamma^{\mathrm{k}}(\tau^*) = \gamma^{\mathrm{k}}(\infty)=\gamma .
\end{equation}
If $\tau^*$ is in the asymptotic region (\ref{eq:gammaklimexp})
one gets in the exponential case (\ref{eq:expformfactor}),
\begin{equation}\label{eq:taujumpexp}
\kappa^{(E)}\left(\frac{\pi}{\tau^*}\right) \simeq
\frac{\pi}{8}\gamma = \frac{\pi^2}{4} \kappa^\beta(\Omega) ,
\end{equation}
which yields
\begin{equation}\label{eq:taustarexp}
\tau^*\simeq -\frac{\pi}{\Lambda}\,
W_{-1}\left(-\frac{\pi}{8}\frac{\gamma}{g^2\Lambda}\right)^{-1}=
-\frac{\pi}{\Lambda}\, W_{-1}\left(-\frac{\pi^2}{4}\frac{
\kappa^\beta(\Omega)}{g^2\Lambda}\right)^{-1} ,
\end{equation}
where $W$ is Lambert's $W$-function \cite{Lambertfun}, that is the
inverse of the function $f(W)=W e^{W}$, and we have taken its $-1$
branch.

\begin{figure}
\begin{center}
\includegraphics[width=\figwidth]{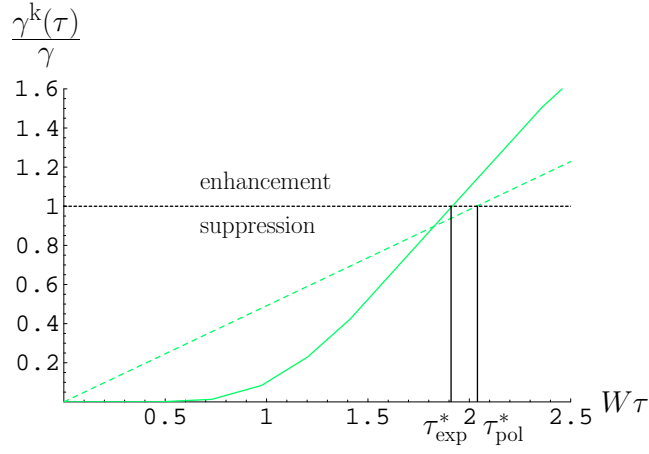}
\end{center}
\caption{BB kicks: $\gamma^{\mathrm{k}}(\tau)/\gamma$ vs $W\tau$
for small $\tau$. Full line: exponential form factor
(\ref{eq:expformfactor}); dashed line: polynomial form factor
(\ref{eq:lorformfactor1}) with $n=2$. Decoherence is suppressed
when $\gamma^{\mathrm{k}}(\tau)<\gamma$; it is enhanced
otherwise.}
\label{fig:kicksmall}
\end{figure}

On the other hand, for the polynomial case (\ref{eq:lorformfactor1}) one
gets from (\ref{eq:gammaklimpow})
\begin{equation}\label{eq:taujumppl}
\kappa^{(P)}\left(\frac{\pi}{\tau^*}\right) \simeq \frac{\pi}{8
\left(1-2^{-2n-1}\right)\zeta(2n+1) }\gamma  =
\frac{\pi^2}{4\left(1-2^{-2n-1}\right)\zeta(2n+1)}
\kappa^\beta(\Omega) ,
\end{equation}
and
\begin{equation}\label{eq:taustarpow}
\tau^*\simeq \frac{\pi}{\Lambda}\,
\left(\frac{\pi}{8\left(1-2^{-2n-1}\right)\zeta(2n+1)}
\frac{\gamma}{g^2\Lambda}\right)^{\frac{1}{2n-1}}=
\frac{3\pi}{\Lambda}\, \left(\frac{\pi^2
}{4\left(1-2^{-2n-1}\right)\zeta(2n+1)}
\frac{\kappa^\beta(\Omega)}{g^2\Lambda}\right)^{\frac{1}{2n-1}} \
.
\end{equation}
The short-time region is shown in Fig.\
\ref{fig:kicksmall}.
It is useful to observe that the results
(\ref{eq:taujumpexp})-(\ref{eq:taustarexp}) and
(\ref{eq:taujumppl})-(\ref{eq:taustarpow}) bear an important
dependence of $\tau^*$ on the ``tail" of the form factor. This is
to be sharply contrasted with the projective measurement situation
(\ref{eq:tjump}), that yields a dependence of the transition time
$\tau^*$ on the ``global" features of the form factor. This
difference is apparent if one compares Figs.\ \ref{fig:meassmall}
and \ref{fig:kicksmall} and shows that the latter method offers
important advantages if one aims at inhibiting decoherence,
because of the larger (and easier to attain) value of $\tau^*$.

\subsection{Control by continuous coupling}

Finally, we can look at continuous coupling. The timescale for
decoherence is (\ref{eq:dressup}):
\begin{eqnarray}
 \gamma^{\mathrm{c}}(K) &=& \pi \int_{-\infty}^\infty d \omega \;
\kappa^\beta(\omega)\;
\left[\delta(\omega-\Omega-K)+\delta(\omega-\Omega+K)\right]
\nonumber\\ &=&
 \pi \left[ \kappa^\beta(\Omega+ K) +
\kappa^\beta(\Omega-K)\right]
 \sim \pi \kappa(K)
 \left(1+e^{-\beta K}\right)\sim \pi \kappa(K), \qquad K\to\infty.
 \label{eq:gammaclim}
\end{eqnarray}
On the other hand,
\begin{eqnarray}
 \gamma^{\mathrm{c}}(K) \to \gamma, \qquad K\to 0 .
\label{eq:gammaclim1}
\end{eqnarray}
Notice that the role of $K$ in Eq.\ (\ref{eq:gammaclim}) and the
role of $1/\tau$ in Eqs.\ (\ref{eq:gammaklimexp}) and
(\ref{eq:gammaklimpow}) are equivalent (see also Appendix
\ref{sec:C}). This yields a natural comparison \cite{bang} between
different timescales ($\tau$ for measurements and kicks, $1/K$ for
continuous coupling).

\begin{figure}
\begin{center}
\includegraphics[width=\figwidth]{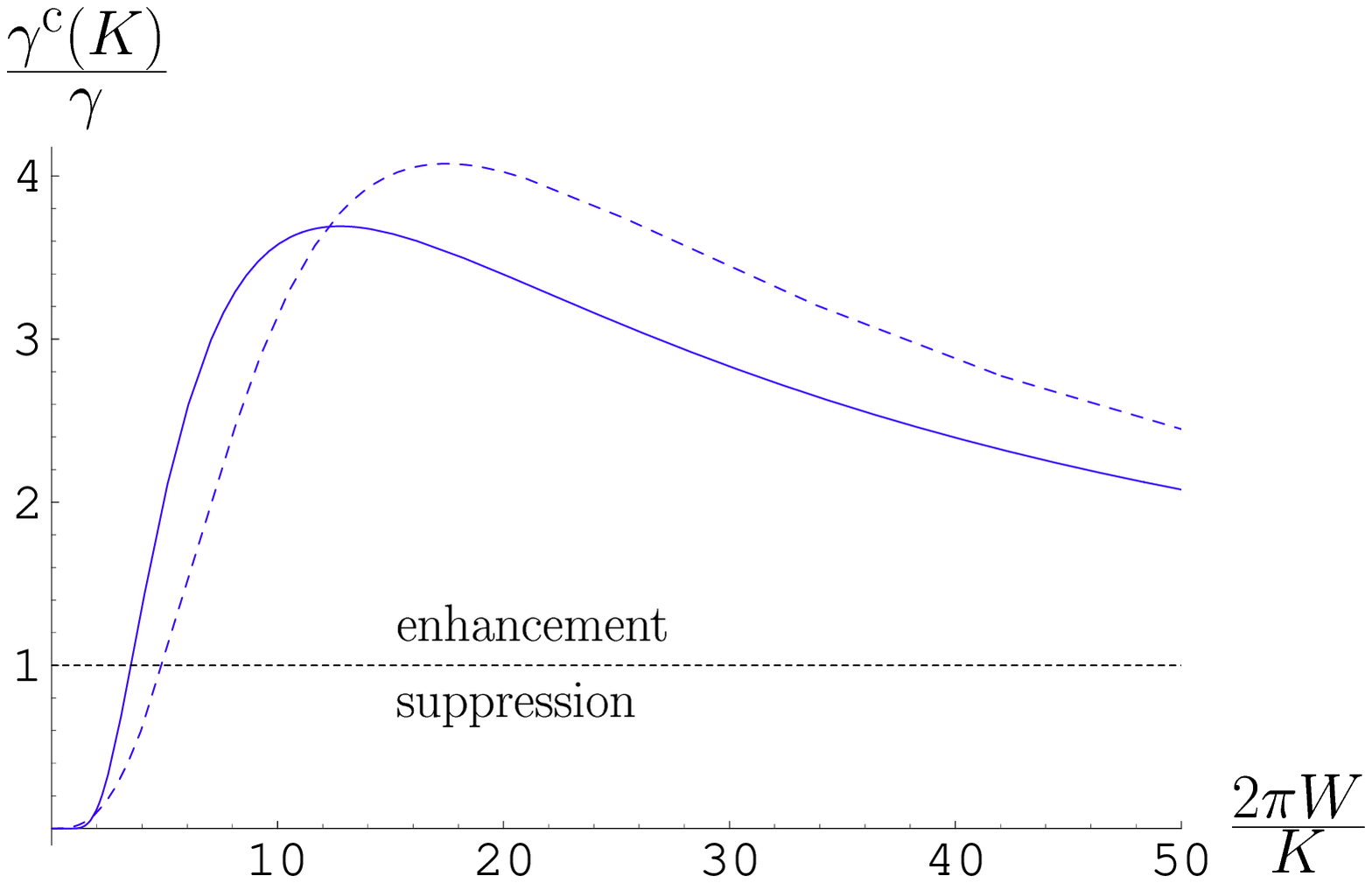}
\end{center}
\caption{Continuous coupling: $\gamma^{\mathrm{c}}(K)/\gamma$ vs
$2\pi W/K$. Full line: exponential form factor
(\ref{eq:expformfactor}); dashed line: polynomial form factor
(\ref{eq:lorformfactor1}) with $n=2$.}
\label{fig:cont}
\end{figure}

\begin{figure}
\begin{center}
\includegraphics[width=\figwidth]{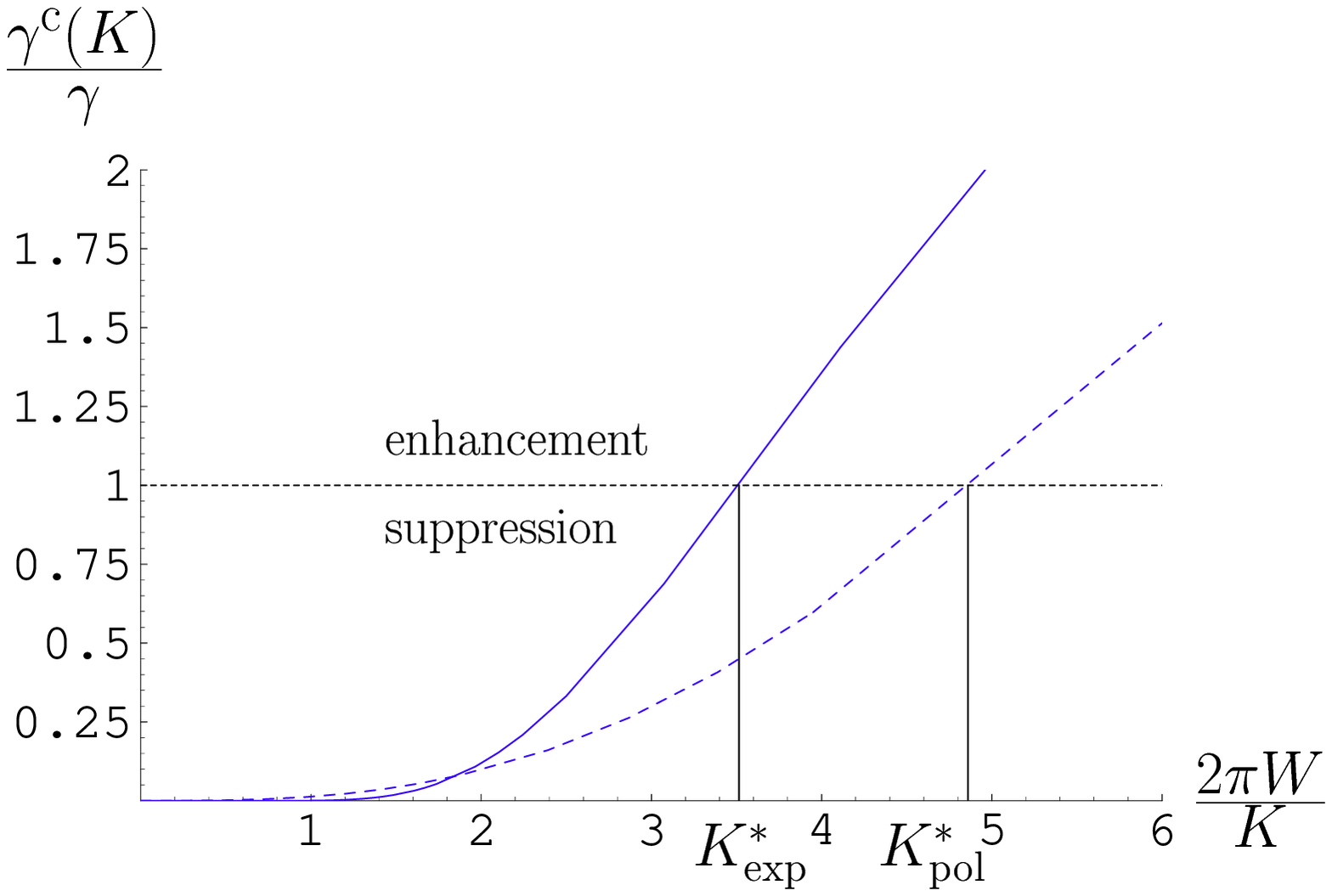}
\end{center}
\caption{Continuous coupling: $\gamma^{\mathrm{c}}(K)/\gamma$ vs
$2\pi W/K$ for large $K$. Full line: exponential form factor
(\ref{eq:expformfactor}); dashed line: polynomial form factor
(\ref{eq:lorformfactor1}) with $n=2$. Decoherence is suppressed
when $\gamma^{\mathrm{c}}(K)<\gamma$; it is enhanced otherwise.}
\label{fig:contsmall}
\end{figure}

The ratio $\gamma^{\mathrm{c}}(K)/\gamma$ is shown in Fig.\
\ref{fig:cont} as a function of $2\pi/K$.
The transition between these two regimes takes now place at
$K=K^*$ where $K^*$ is defined by the equation
\begin{equation}\label{eq:K*def}
 \gamma^{\mathrm{c}}(K^*) = \gamma^{\mathrm{c}}(0)=\gamma .
\end{equation}
If $K^*$ is in the asymptotic region (\ref{eq:gammaclim})
\begin{equation}\label{eq:Kjump}
\kappa(K^*) \simeq \frac{\gamma}{\pi} = 2 \kappa^\beta(\Omega) .
\end{equation}
For the exponential form factor (\ref{eq:expformfactor}) one gets
\begin{equation}\label{eq:Kstarexp}
K^*\simeq -\Lambda\, W_{-1}\left(-\frac{1}{\pi}\frac{\gamma}{
g^2\Lambda}\right)= -\Lambda\, W_{-1}\left(-2\frac{
\kappa^\beta(\Omega)}{g^2\Lambda}\right) ,
\end{equation}
while for the polynomial form factor (\ref{eq:lorformfactor1}) one
gets
\begin{equation}\label{eq:Kstarpow}
K^*\simeq \Lambda\, \left(\frac{1}{\pi}\frac{\gamma}{
g^2\Lambda}\right)^{-\frac{1}{2n-1}}= \Lambda\, \left(2\frac{
\kappa^\beta(\Omega)}{g^2\Lambda}\right)^{-\frac{1}{2n-1}} \ .
\end{equation}
One observes a dependence of $K^*$ on the tail of the form factor.
The strong coupling region is shown in Fig.\ \ref{fig:contsmall}.

\subsection{Comparison among the three control strategies}

\begin{figure}
\begin{center}
\includegraphics[width=0.9\textwidth]{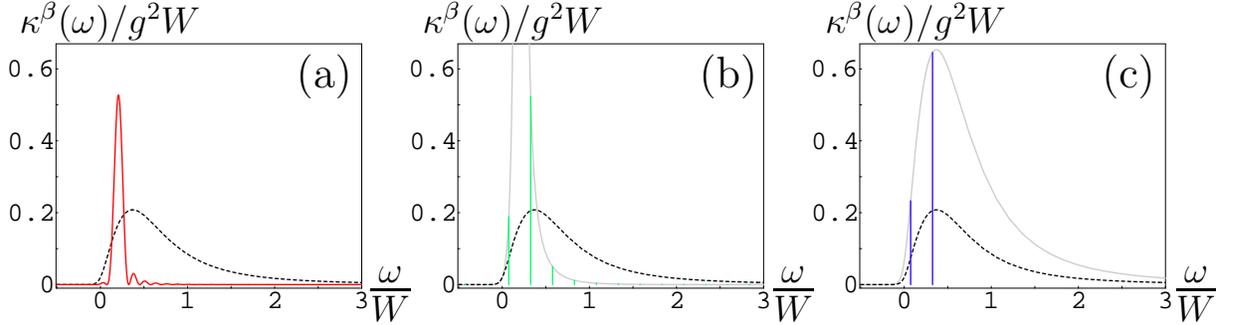}
\end{center}
\caption{Different features of the three control methods. Form
factor (polynomial, $n=2$) $\kappa^\beta(\omega)$ (dashed line)
and form factor modulated/multiplied by the control ``response"
function (full line) for: (a) pulsed measurements, Eq.\
(\ref{eq:gammazenolim}), with control response function $\tau{\rm
sinc}^2[(\omega-\Omega)\tau/2]$ (here and in the other two
cases, $\Omega=0.2W$); (b) BB kicks, Eq.\ (\ref{eq:gammaklim}),
with control response function $(2/\pi)\sum_{j=0}^\infty
\left(j+\frac{1}{2}\right)^{-2}
 \left[\delta\left(\omega-\Omega-\frac{\pi}{\tau}(2j+1)\right)
+\delta\left(\omega-\Omega+\frac{\pi}{\tau}(2j+1)\right)\right]$
[see (\ref{eq:intkick}) and notice that the first 2-3 terms of the
series yield an excellent approximation]; (c) continuous
measurement, Eq.\ (\ref{eq:gammaclim}), with control response
function $\pi[\delta(\omega-\Omega-K)+\delta(\omega-\Omega+K)]$.
The gray line is a guide for the eye and interpolates $(2/\pi)
\left(j+\frac{1}{2}\right)^{-2} \kappa^\beta(\omega)$ in (b) and
$\pi\kappa^\beta(\omega)$ in (c). We set $\tau=2\pi /K= 50 W^{-1}$
(a ``large" value): this yields in all cases a (controlled) decay
rate that is very close to that obtained by the Fermi Golden rule.
}
\label{fig:differenceFGR}
\end{figure}

\begin{figure}
\begin{center}
\includegraphics[width=0.9\textwidth]{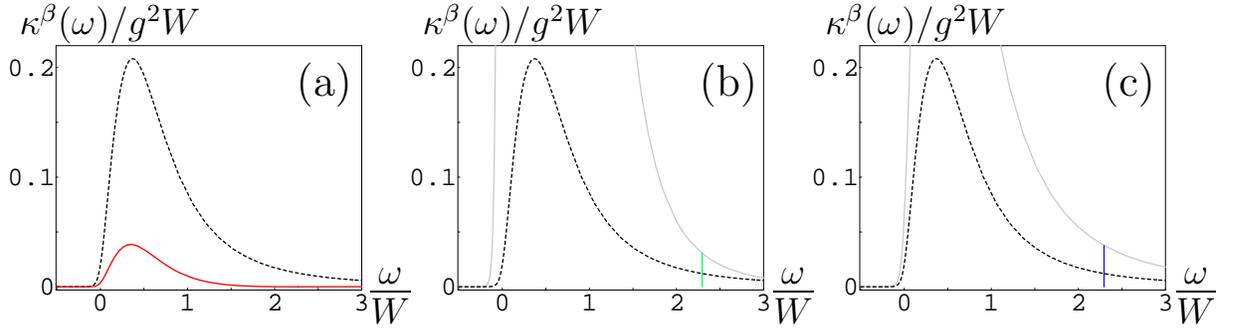}
\end{center}
\caption{Same as in Fig.\ \ref{fig:differenceFGR}, but for
$\tau=2\pi/K=3 W^{-1}$ (a ``small" value): this yields a
\emph{bona fide} control of the decay rate (in this particular
situation, decoherence is enhanced in the Zeno case and suppressed
in the other two cases). (a) The control response function
$\tau{\rm sinc}^2[(\omega-\Omega)\tau/2]$ is very broad and the
effective lifetime depends on the ``global" features of the form
factor. (b) For small $\tau$ all the arguments of the
$\delta$-functions in (\ref{eq:intkick}) tend to $\infty$: for
well-behaved form factors (like that shown in the figure), only
the \emph{first} term contributes significantly; the controlled
lifetime depends on the local features of the ``tail" of the form
factor. (c) For large $K$ the arguments of the $\delta$-functions
in Eq.\ (\ref{eq:gammaclim}) tend to $\pm \infty$ and the
controlled lifetime depends again on the local features of the
``tail" of the form factor. }
\label{fig:differenceQZE}
\end{figure}

There is a clear difference between \emph{bona fide} projective
measurements and the other two cases, BB kicks and continuous
coupling. In the former case Eqs.\
(\ref{eq:tau*def})-(\ref{eq:tjump}) yield a dependence of $\tau^*$
on the global features of the form factor (i.e., its integral). By
contrast, Eqs.\ (\ref{eq:taujumpexp})-(\ref{eq:taustarpow}) and
(\ref{eq:Kjump})-(\ref{eq:Kstarpow}) ``pick" some particular
(``on-shell") value(s). This important difference, due to the
different features of the evolution (non-unitary in the first
case, unitary in the latter cases), is graphically displayed in
Fig.\ \ref{fig:differenceFGR} and \ref{fig:differenceQZE}, where
the different mechanisms of control are compared. In Fig.\
\ref{fig:differenceFGR}, $\tau$ is ``large" (in units of inverse
bandwidth) and the three methods yield almost \emph{no} control:
one essentially reobtains the Fermi Golden rule $\gamma = 2 \pi
\kappa^\beta(\Omega)$, although in different ways. In Fig.\
\ref{fig:differenceQZE}, $\tau$ is ``small" and the effective
lifetime is sensibly modified, although by different mechanisms.

\begin{figure}
\begin{center}
\includegraphics[width=\figwidth]{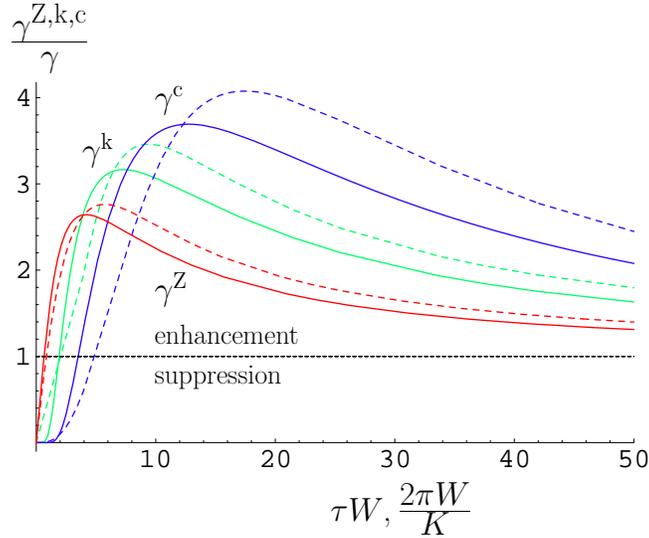}
\end{center}
\caption{Comparison among the three control methods. The graphs of
Figs.\ \ref{fig:meas}, \ref{fig:kick} and \ref{fig:cont} are
displayed together. BB kicks and continuous coupling are more
effective than \emph{bona fide} measurements for combatting
decoherence, as the regime of ``suppression" is reached for larger
values of $\tau$ and $K^{-1}$. } \label{fig:confr}
\end{figure}

\begin{figure}
\begin{center}
\includegraphics[width=\figwidth]{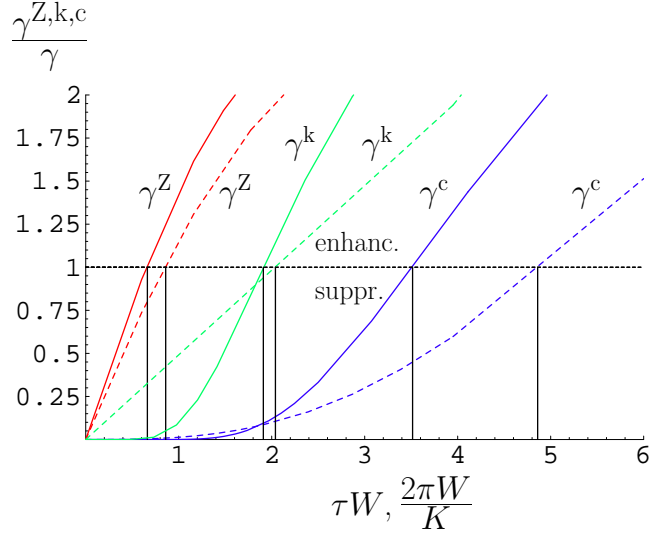}
\end{center}
\caption{Comparison among the three control methods:
small times/strong coupling regions. The graphs of Figs.\
\ref{fig:meassmall}, \ref{fig:kicksmall} and \ref{fig:contsmall}
are displayed together.}
\label{fig:confrsmall}
\end{figure}

The three control methods are graphically compared in Figs.\
\ref{fig:confr}-\ref{fig:confrsmall}. The different features
discussed in Figs.\
\ref{fig:differenceFGR}-\ref{fig:differenceQZE} yield very
different outputs, clearly apparent in Fig.\ \ref{fig:confrsmall},
that can be important in practical applications: decoherence can
be more easily halted by applying BB and/or continuous coupling
strategies. These two methods yield values of $\tau^*$ (or $K^*$)
that are easier to attain. However, this advantage has a price,
because BB and continuous coupling yield a larger enhancement of
decoherence for $\tau
>\tau^*$, $K < K^*$. The two dynamical methods perform better only
when $\tau \lesssim \tau^*$, $K \gtrsim K^*$. This is apparent in
Fig.\ \ref{fig:confr}. We notice that a strict comparison between
continuous coupling and the other two methods is difficult, as it
would involve an analysis of numerical factors of order one in the
definition of the relevant conversion factors between the
frequency of interruptions $\tau$ and the coupling $K$ [this
factor has been sensibly--but arbitrarily--set equal to $2\pi$ in
Figs.\ \ref{fig:confr}-\ref{fig:confrsmall}: see sentence after
Eq.\ (\ref{eq:gammaclim1}) and Appendix \ref{sec:C}].


\section{SUMMARY AND CONCLUDING REMARKS}
\label{sec.conclusions}

We have analyzed and compared three control methods for combatting
decoherence. The first is based on repeated quantum measurements
(projection operators) and involves a description in terms of
nonunitary processes. The second and third methods are both
dynamical, as they can be described in terms of unitary
evolutions. In all cases, decoherence can be halted by very
rapidly/strongly driving or very frequently measuring the system
state. However, if the frequency is not high enough or the
coupling not strong enough, the controls may accelerate the
decoherence process and deteriorate the performance of the quantum
state manipulation. The acceleration of decoherence is analogous
to the inverse Zeno effect, namely the acceleration of the decay
of an unstable state due to frequent measurements
\cite{IZE,Heraclitus}.

As a general rule, when one endeavors to control decoherence by
suitably tailoring the coupling of the system of interest to
another system (such as an external field, or a measuring
apparatus), one should carefully look at the relevant timescales,
as it is not true that repeated measurements/interruptions always
lead to a suppression of decoherence.

It is convenient to summarize the main results obtained in this
article in the particular case of a two-level system (qubit) with
energy difference $\Omega$. If the frequency $\tau^{-1}$ of
measurements or BB kicks, or the strength $K$ of the coupling tend
to $\infty$, the two-dimensional (Zeno) subspace defining the
qubit becomes isolated and decoherence is completely suppressed.
However, if $\tau^{-1}$ and $K$ are large, but not extremely
large, the transition (decay) rates between the qubit subspace and
the remaining sector of the Hilbert space display a complicated
dependence on $\tau^{-1}$ and $K$, and decoherence can be
suppressed or enhanced, depending on the situation.

At low temperature $\beta^{-1} \ll \Omega \ll W$, where $W$ is the
bandwidth of the form factor of the interaction, the decay rates
read, from (\ref{eq:gammazenolim}), (\ref{eq:tauZtherm}),
(\ref{eq:gammaklimexp}), (\ref{eq:gammaklimpow}) and
(\ref{eq:gammaclim})
\begin{equation}
\left\{
\begin{array}{l}
\gamma^Z(\tau) \sim \frac{\tau}{\tau_Z^2}, \qquad \tau\to0
 \ , \\
\\
\gamma^{\mathrm{k}}(\tau) \sim \frac{8}{\pi}
\kappa\left(\frac{\pi}{\tau}\right), \qquad \tau\to0 \ , \\
\\
\gamma^{\mathrm{c}}(K) \sim \pi \kappa(K), \qquad K\to\infty \ ,
\end{array}
\right. \label{eq:gammasimp}
\end{equation}
where $Z$, k and c denote (Zeno) measurements, (BB) kicks and
continuous coupling, respectively, $\kappa$ is the form factor and
$ 1/\tau_Z^2 \simeq\int d\omega \kappa(\omega)$ the Zeno time
(more accurate definitions were given in the preceding sections).
As we have shown, there is a characteristic transition time
$\tau^*$ [coupling $K^*$], such that one obtains:
\begin{eqnarray}
\mbox{decoherence suppression}: & &\gamma(\tau)<\gamma \quad
[\gamma(K)<\gamma],\qquad \mbox{for} \quad \tau<\tau^* \quad
[K>K^*],
\nonumber\\
\mbox{decoherence enhancement}: & &\gamma(\tau)>\gamma \quad
[\gamma(K)>\gamma],\qquad \mbox{for} \quad \tau>\tau^* \quad
[K<K^*].
\end{eqnarray}
 Therefore, in order to obtain a suppression of
decoherence, the interruptions/coupling must be \emph{very}
frequent/strong. Notice, in this context, that both $\tau^*$ and
$2\pi/ K^*$ are not simply related to the inverse bandwidth $2\pi
W^{-1}$: they can be in general (much) shorter. For instance, in
the Ohmic polynomial  case (\ref{eq:lorformfactor1}), one easily
gets from (\ref{eq:BW}) and (\ref{eq:gammasimp})
\begin{equation}
\left\{
\begin{array}{l}
\tau^*_{Z}\simeq 2\pi W^{-1}
\left(2(n-1)\alpha_n^2\frac{\Omega}{W}\right)
 \ll 2\pi W^{-1} \ , \\
\\
\tau^*_{\mathrm{k}}\simeq 2\pi W^{-1} \frac{\alpha_n}{2}
\left(\frac{\alpha_n\pi^2}{4}\frac{\Omega}{W}\right)^{\frac{1}{2n-1}}
 \ll 2\pi W^{-1}\ ,\\
\\
K^*\simeq W
\alpha_n^{-1}\left(\frac{2}{\alpha_n}\frac{W}{\Omega}\right)^{\frac{1}{2n-1}}
 \gg W \ ,
\end{array}
\right. \label{eq:izlimit}
\end{equation}
where $\alpha_n=(\sqrt{\pi}/2)\Gamma(n-3/2)/\Gamma(n-1)\leq \pi/2$
is a coefficient of order $1$ and $n$ characterizes the polynomial
fall off of the form factor (\ref{eq:lorformfactor1}). The above
times/coupling may be (very) difficult to achieve in practice. In
fact, we see here that the relevant timescale is not simply the
inverse bandwidth $2\pi W^{-1}$, but can be much shorter if
$\Omega \ll W$, as is typically the case. These conclusions,
summarized here for the simple case of a qubit, are valid in
general, when one aims at protecting from decoherence an
$N$-dimensional Hilbert subspace.

An important example that we have not explicitly analyzed in this
article is the case of $1/f$ noise, and its suppression by means
of techniques like those discussed here. There has recently been a
surge of interest in this issue in quantum information processing
devices, where such noise is often attributable to (but certainly
not limited to) charge fluctuations in electrodes providing
control voltages \cite{Galperin:03,Makhlin:03}. The need for such
electrodes is widespread in quantum computer proposals, e.g.,
trapped ions (where observed $1/f$ noise was reported in
\cite{Turchette:00}), quantum dots \cite{Burkard:99}, doped
silicon \cite{Kane:98,Vrijen:00}, electrons on helium
\cite{Platzman:99}, and superconducting qubits \cite{Paladino:02}.
In the latter case, in a recent experiment involving a charge
qubit in a small superconducting electrode (Cooper-pair box), a
spin-echo-type version of BB was successfully used to suppress
low-frequency energy-level fluctuations (causing dephasing) due to
$1/f$ charge noise \cite{Nakamura:02}. Several recent papers have
dealt with suppression of this particular kind of noise via BB
decoupling \cite{ShiokawaLidar,Gutmann:03,Faoro:03,Falci:03}. The
``bottom-up" approach models $1/f$ noise as arising from a
collection of bi-stable fluctuators
\cite{Gutmann:03,Faoro:03,Falci:03,Galperin:03}. The alternative
is to treat $1/f$ noise as contributing a particular form factor
\cite{ShiokawaLidar,Makhlin:03}. We will pursue these ideas as a
future topic of investigation, but we expect that the main results
obtained in the present paper be applicable to this case as well.

The results obtained in this paper are of general validity and
bring to light the different features of the control procedures as
well as the crucial role played by the form factor of the
interaction. We do not expect any drastic change for different
decoherence mechanisms and/or different physical systems. The only
somewhat delicate issue, in our opinion, is to understand whether
the system investigated can be consistently described by means of
a set of discrete levels.

\acknowledgments
This work is partly  supported by the bilateral Italian-Japanese
project 15C1 on ``Quantum Information and Computation'' of the
Italian Ministry for Foreign Affairs, by a Grant-in-Aid for
Scientific Research (C) from JSPS, by a Grant-in-Aid for
Scientific Research of Priority Areas ``Control of Molecules in
Intense Laser Fields'' and the 21st Century COE Program at Waseda
University ``Holistic Research and Education Center for Physics of
Self-organization Systems'' both from the Ministry of Education,
Culture, Sports, Science and Technology of Japan. D.A.L.\
gratefully acknowledges financial support from NSERC, the Sloan
Foundation, and the DARPA-QuIST program (managed by AFOSR under
agreement No. F49620-01-1-0468).

\appendix
\section{}
\label{sec:A}
In this Appendix, the assumption of the factorized form of the
initial density operator, as in~(\ref{eq:rhoprod}), which is
usually taken for granted, is shown to be justified in the
weak-coupling (scaling) limit. We only outline the main
derivation. Further details will be reported elsewhere
\cite{factorassumption}.

Consider the initial-value problem
\begin{equation}
{\partial\over\partial t}\rho
=\cL_{\mathrm{tot}}\rho=(\cL_0+\lambda\cL_{SB})\rho
=(\cL_S+\cL_B+\lambda\cL_{SB})\rho,\quad\rho(0)=\rho_0,
\label{eq:A1}
\end{equation}
where the dependence on the coupling constant $\lambda$ of the
interaction Liouvillian $\cL_{SB}$ is made explicit. Notice that
the initial density operator can be of any form and is {\it not\/}
assumed here to be factorized like in (\ref{eq:rhoprod}). The
projection operators $\mathcal P$ and $\mathcal Q$, defined in
(\ref{eq:defproj}), and the above Liouvillians $\cL_S,\,\cL_B$ and
$\cL_{SB}$ satisfy the same conditions (\ref{eq:propPQ}) and
(\ref{eq:propPQ1}). The projected density operators $\mathcal
P\rho$ and $\mathcal Q\rho$ satisfy
\begin{equation}
{\partial\over\partial t}{\mathcal P}\rho =\cL_0{\mathcal
P}\rho+\lambda{\mathcal P}\cL_{SB}{\mathcal Q}\rho,\qquad
{\partial\over\partial t}{\mathcal Q}\rho =(\cL_0+\lambda{\mathcal
Q}\cL_{SB}{\mathcal Q}){\mathcal Q}\rho
  +\lambda{\mathcal Q}\cL_{SB}{\mathcal P}\rho,
\label{eq:A3}
\end{equation}
respectively. Following the same procedure as in
Sec.~\ref{sec.model}, we arrive at the following {\it exact\/}
equation for the ${\mathcal P}$-projected operator in the
interaction picture
\begin{eqnarray}
{\partial\over\partial t}\left(e^{-\cL_0t}{\mathcal P}\rho\right)
&=&\lambda e^{-\cL_0t}{\mathcal P}\cL_{SB}
 e^{(\cL_0+\lambda{\mathcal Q}\cL_{SB}{\mathcal Q})t}{\mathcal Q}\rho_0
 \nonumber\\
  & &+
 \lambda^2\int_0^tdt'e^{-\cL_0t}{\mathcal P}\cL_{SB}
 e^{(\cL_0+\lambda{\mathcal Q}\cL_{SB}{\mathcal Q})t'}\cL_{SB}{\mathcal P}\rho(t-t').
\label{eq:A4}
\end{eqnarray}

Notice that the first term on the rhs represents the contribution
arising from a possible initial correlation between the system and
reservoir. We now show that this term dies out in the
weak-coupling (i.e., scaling) limit $\lambda\to0$ with fixed
$\tau\equiv\lambda^2t$. For this purpose, define
\begin{equation}
\rho_{\mathrm{I}}(\tau;\lambda)\equiv
e^{-\cL_0\tau/\lambda^2}{\mathcal P}\rho(\tau/\lambda^2),
\label{eq:A5}
\end{equation}
that satisfies
\begin{eqnarray}
\dot\rho_{\mathrm{I}}(\tau;\lambda)
&=&{1\over\lambda}e^{-\cL_0\tau/\lambda^2}{\mathcal P}\cL_{SB}
 e^{(\cL_0+\lambda{\mathcal Q}\cL_{SB}{\mathcal Q})\tau/\lambda^2}{\mathcal Q}\rho_0\nonumber\\
 & &+
 \int_0^{\tau/\lambda^2}dt'e^{-\cL_0\tau/\lambda^2}{\mathcal P}\cL_{SB}
 e^{(\cL_0+\lambda{\mathcal Q}\cL_{SB}{\mathcal Q})t'}\cL_{SB}
 e^{\cL_0(\tau/\lambda^2-t')}{\mathcal P}
 \rho_{\mathrm{I}}(\tau-\lambda^2t';\lambda).
\label{eq:A6}
\end{eqnarray}
The first term vanishes in the $\lambda\to0$ limit
\cite{factorassumption}, since
\begin{equation}
\int_0^\infty d\tau{1\over\lambda}e^{\mathcal
A\tau/\lambda^2}Y(\tau) =\lambda\int_0^\infty d\tau e^{\mathcal
A\tau}Y(\lambda^2\tau) \longrightarrow0,\quad\hbox{as
}\quad\lambda\to0,
\label{eq:A7}
\end{equation}
for any superoperator such that the integral $\int_0^\infty d\tau
e^{\mathcal A\tau}$ exists. This means that the contribution
originating from the initial correlation between the system and
reservoir disappears in the scaling limit and therefore we are
allowed to start from an initial density matrix in the factorized
form (\ref{eq:rhoprod}).

Finally, the dynamics of $\rho_{\mathrm{I}}(\tau;0)$ is governed
by
\begin{equation}
\dot\rho_{\mathrm{I}}(\tau;0) =\sum_\omega\tilde
Q_\omega\int_0^\infty dt'{\mathcal P}\cL_{SB}(0) {\mathcal
Q}\cL_{SB}(-t')\tilde Q_\omega\rho_{\mathrm{I}}(\tau;0)
\label{eq:A8}
\end{equation}
with the factorized initial condition (\ref{eq:rhoprod}), where
the  $\tilde Q_\omega$ are the eigenprojections of the Liouvillian
$\cL_S$ defined in (\ref{eq:eigenA9}).

{}From a physical point of view, the factorization ansatz
described in this appendix simply means that the ``initial"
correlations between the system and its environment are
``forgotten" on a time scale of order $\lambda^2$. We also note
that several authors have addressed the question of the
modifications that arise when it is not permissible to assume
initially separable system-environment, e.g., \cite{factorvari}.

\section{}
\label{sec:B}

We derive Eq.\ (\ref{ZenoApp}). The first equality reads
\begin{eqnarray}
\left[{\hat P} e^{\cL_{\rm tot} \tau} {\hat P}\right]^{t/\tau}
\simeq \left[{\hat P} V_Z(\tau){\hat P}\right]^{t/\tau} , \qquad
V_Z(\tau)=e^{\cL_0 \tau} {\cal T} \exp\left(\int_0^{\tau} ds\;
e^{-\cL_0 s} \cG_{\mathrm{Z}}(\tau) e^{\cL_0 s}\right).
\label{ZenoAppB0}
\end{eqnarray}
Let us write $V_Z(\tau)=V(\tau,\tau)$, where
\begin{equation}\label{eq:Vtudef}
V(t,u)=e^{\cL_0 t} {\cal T} \exp\left(\int_0^{t} ds\; e^{-\cL_0 s}
\cG_{\mathrm{Z}}(u) e^{\cL_0 s}\right).
\end{equation}
By deriving with respect to $t$, we get
\begin{equation}\label{eq:Vtudef1}
\partial_tV(t,u)=[\cL_0 + \cG_{\mathrm{Z}}(u)] V(t,u),
\end{equation}
so that
\begin{equation}\label{eq:Vtudef2}
V(t,u)= \exp\{[\cL_0 + \cG_{\mathrm{Z}}(u)]t \} ,
\end{equation}
where we used $V(0,u)=1$. As a consequence,
$V_Z(\tau)=\exp\{[\cL_0 + \cG_{\mathrm{Z}}(\tau)]\tau \}$ and
\begin{eqnarray}
\left[{\hat P} e^{\cL_{\rm tot} \tau} {\hat P}\right]^{t/\tau}
\simeq \left[{\hat P}\exp\{[\cL_0 + \cG_{\mathrm{Z}}(\tau)]\tau \}
{\hat P}\right]^{t/\tau} = {\hat P}\exp\{[\cL_0 +
\cG_{\mathrm{Z}}(\tau)]t \} , \label{ZenoAppB1}
\end{eqnarray}
because $[\hat P,\cL_0]=[\hat P,\cG_{\mathrm{Z}}]=0$. This is Eq.\
(\ref{ZenoApp}).

Let us now solve Eq.\ (\ref{cLZ}):
\begin{equation}\label{eq:a1cLZ}
\int_0^\tau dt\; e^{-\cL_S t}\cL_{\mathrm{Z}}(\tau)e^{\cL_S t}=
\int_0^\tau dt {\hat P} \cL_{\mathrm{I}}(t){\hat P}= \int_0^\tau
dt\int_0^t ds\; {\hat P}\cK_{\mathrm{I}}(t,s){\hat P} \ .
\end{equation}
By using (\ref{eq:cKI}) and (\ref{eq:propnew})
\begin{equation}\label{eq:a1cKI}
\cK_{\mathrm{I}}(t,s)\, \sigma = \mathrm{tr}_B
\left\{\cL_{SB}(t)\cL_{SB}(s)\, \sigma\otimes\rho_B \right\},
\qquad \cL_{SB}(t)= e^{-(\cL_S+\cL_B)t} \cL_{SB}
e^{(\cL_S+\cL_B)t} ,
\end{equation}
we get
\begin{eqnarray}\label{eq:a1cLZ1}
\int_0^\tau dt\; e^{-\cL_S t}\cL_{\mathrm{Z}}(\tau)e^{\cL_S t}
\sigma &=& \int_0^\tau dt\int_0^t ds\;  \mathrm{tr}_B \left\{{\hat
P}\cL_{SB}(t)\cL_{SB}(s)\, {\hat P}\sigma\otimes\rho_B \right\}
\nonumber\\
& = & \int_0^\tau dt \; e^{-\cL_S t}   \int_{-t}^0
ds\;\mathrm{tr}_B \left\{{\hat P}\cL_{SB}\cL_{SB}(s) {\hat
P}e^{\cL_S t}\sigma\otimes\rho_B \right\} .
\end{eqnarray}
Let us rewrite the previous equation in terms of the
eigenprojections $\tilde Q_\omega$ of $\cL_S$ defined by
(\ref{eq:eigenA9}):
\begin{equation}\label{eq:pwa1cKI}
\sum_{\omega,\omega'} \int_0^\tau dt\; e^{i(\omega-\omega')t}
\tilde Q_\omega\cL_{\mathrm{Z}}(\tau)\tilde Q_{\omega'} \sigma =
\sum_{\omega,\omega'} \int_0^\tau dt \; e^{i(\omega-\omega')t}
\int_{-t}^0 ds\;\mathrm{tr}_B \left\{\tilde Q_\omega{\hat
P}\cL_{SB}\cL_{SB}(s) {\hat P}\tilde
Q_{\omega'}\sigma\otimes\rho_B \right\}.
\end{equation}
Performing the first integral, we get
\begin{equation}\label{eq:pwa1cKI2}
\tilde Q_\omega\cL_{\mathrm{Z}}(\tau)\tilde Q_{\omega'} \sigma =
\frac{g((\omega-\omega')\tau)}{\tau} \int_0^\tau dt \;
e^{i(\omega-\omega')t} \int_{-t}^0 ds\;\mathrm{tr}_B \left\{\tilde
Q_\omega{\hat P}\cL_{SB}\cL_{SB}(s) {\hat P}\tilde
Q_{\omega'}\sigma\otimes\rho_B \right\}, \qquad
g(x)=\frac{ix}{e^{ix}-1} .
\end{equation}
Since $g(0)=1$, the diagonal terms yield
\begin{equation}\label{eq:diagpwa1cKI2}
\tilde Q_\omega\cL_{\mathrm{Z}}(\tau)\tilde Q_{\omega} \sigma =
\frac{1}{\tau} \int_0^\tau dt  \int_{-t}^0 ds\;\mathrm{tr}_B
\left\{\tilde Q_\omega{\hat P}\cL_{SB}\cL_{SB}(s) {\hat P}\tilde
Q_{\omega}\sigma\otimes\rho_B \right\} .
\end{equation}
The off-diagonal terms do not contribute to the master equation,
as explained at the end of Sec.\ \ref{sec.model}, Eqs.\
(\ref{eq:formal})-(\ref{eq:P2ndorder}).

By using the property (\ref{eq:prl1l2}) and noting that $[\hat P,
\tilde Q_\omega]=0$ by (\ref{eq:nozeno}), we get
\begin{equation}\label{eq:diagpwa1cKI3}
\cL_{\mathrm{Z}}(\tau) \sigma = -\sum_\omega \frac{1}{\tau}
\int_0^\tau dt \int_{-t}^0 ds\; \mathrm{tr}_B \left\{{\hat P}
\left[\left(\tilde Q_\omega H_{SB}\right),\left[\left(\tilde
Q_{-\omega} H_{SB}(s)\right),{\hat
P}\sigma\otimes\rho_B\right]\right] \right\} ,
\end{equation}
whence, by using (\ref{eq:pwHsb}),
\begin{eqnarray}
\cL_{\mathrm{Z}}(\tau) \sigma &=& -\sum_m \frac{1}{\tau}
\int_0^\tau dt \int_{-t}^0 ds\; \mathrm{tr}_B \left\{{\hat P}
\left[H_{SB}^{(m)},\left[ H_{SB}^{(-m)}(s),{\hat
P}\sigma\otimes\rho_B\right]\right] \right\} \nonumber\\
& = & -\sum_m \frac{1}{\tau} \int_0^\tau dt \int_{-t}^0 ds\;
\mathrm{tr}_B \left\{ {\hat P} \left[X^\dagger_m \otimes
A_m,\left[ X_{m} \otimes A^\dagger_{m}(s),{\hat
P}\sigma\otimes\rho_B\right]\right]
\right. \nonumber\\
& &\qquad\qquad\qquad\qquad\qquad\quad \left. + {\hat P}
\left[X_{-m} \otimes A^\dagger_{-m},\left[ X^\dagger_{-m} \otimes
A_{-m}(s),{\hat P}\sigma\otimes\rho_B\right]\right]\right\},
\label{eq:diagpwa1cKIm}
\end{eqnarray}
where, like in Eq.\ (\ref{eq:diagcL}), in the second equality we
neglected terms containing two annihilation or creation operators.
From (\ref{eq:diagpwa1cKIm}) we get Eq.\ (\ref{eq:disscLz}) with
\begin{equation}\label{eq:gammazenodefa1}
\gamma_{m}^{\mathrm{Z}}(\tau)= \frac{2}{\tau} \Re \int_0^\tau dt
\int_{-t}^0 ds \left(\left\langle A_m(0)
A^\dagger_m(s)\right\rangle+ \left\langle A_{-m}^\dagger(0)
A_{-m}(s)\right\rangle\right) .
\end{equation}
By noticing that
\begin{equation}\label{eq:lrlr}
\left\langle A_m A^\dagger_m(s)\right\rangle+ \left\langle
A_{-m}^\dagger A_{-m}(s)\right\rangle=\int_{-\infty}^\infty
d\omega\; \kappa^\beta_m(\omega) e^{i(\omega-\omega_m) s},
\end{equation}
we finally get
\begin{equation}\label{eq:gammazenodefa2}
\gamma_{m}^{\mathrm{Z}}(\tau)
=\frac{2}{\tau}\int_{-\infty}^\infty d\omega\;
\kappa^\beta_m(\omega)
\frac{1-\cos(\omega-\omega_m)\tau}{(\omega-\omega_m)^2}=
\tau\int_{-\infty}^\infty d\omega\; \kappa^\beta_m(\omega)
\frac{\sin^2\left(\frac{\omega-\omega_m}{2}\tau\right)}{\left(\frac{\omega-
\omega_m}{2}\tau\right)^2},
\end{equation}
which is Eq.\ (\ref{eq:gammazenodef}) of the text.

\section{}
\label{sec:C}
We derive Eqs.\ (\ref{eq:disscLk}) and (\ref{eq:gammakick}). We
start from Eqs.\ (\ref{noname1}) and (\ref{noname2})
\begin{eqnarray}
& & \int_0^{\tau} ds\; e^{-\cL_\tau s} \cF_\mathrm{k} (\tau)
e^{\cL_\tau s}= \int_0^\tau ds\; \cL_{SB}(s) \ ,
\\
& & \int_0^{\tau} ds\; e^{-\cL_\tau s} \cG_\mathrm{k} (\tau)
e^{\cL_\tau s}= \int_0^\tau ds\int_0^s ds_1\;
\left[\cL_{SB}(s)\cL_{SB}(s_1)- e^{-\cL_\tau s}\cF_\mathrm{k}
(\tau) e^{\cL_\tau (s-s_1)}\cF_\mathrm{k} (\tau) e^{\cL_\tau
s_1}\right] \ ,
\end{eqnarray}
where $\cL_\tau=\frac{\cL_\mathrm{k}}{\tau}+\cL_0$, and by taking
the trace over the bath we get
\begin{eqnarray}\label{eq:a1cLk1}
\int_0^\tau dt\; e^{-\cL'_S t}\cL_{k}(\tau)e^{\cL'_S t} \sigma =
\int_0^\tau dt    \int_{-t}^0 ds\;\mathrm{tr}_B \left\{
\left[e^{-\cL_S t} \cL_{SB}\cL_{SB}(s)- e^{-\cL'_S
t}\cF_\mathrm{k} (\tau) e^{-\cL_\tau s}\cF_\mathrm{k} (\tau)
e^{\cL_\tau s} e^{\cL'_S t}\right]\sigma\otimes\rho_B \right\} ,
\end{eqnarray}
with
\begin{equation}
\cL_S (\tau)=\frac{\cL_{\mathrm{k}}}{\tau}+\cL_S .
\end{equation}
Equation (\ref{eq:a1cLk1}) is similar to (\ref{eq:a1cLZ1}) and, by
projecting onto the eigenprojections $\tilde P_\omega(\tau)$ of
$\cL_S(\tau)$ and taking only the diagonal terms, one obtains
Eq.~(\ref{eq:disscLk}). However, in order to compute the decay
rates $\gamma_m^{\mathrm{k}}(\tau)$ one can give an alternative,
more physical derivation by elaborating on the technique of Ref.\
\cite{bang}. First notice that the BB dynamics (\ref{noname0}) is
generated by the time-dependent Hamiltonian
\begin{equation}\label{eq:Hamk}
H\left(t/\tau\right)= H_{\mathrm{tot}} + H_{\mathrm{k}}\;
\delta_P\left(t/\tau \right), \quad
\delta_P(x)=\sum_{n\in\mathbb{Z}} \delta(x-n) \ .
\end{equation}
In the enlarged Hilbert space $\mathcal{H}\otimes L^2(\mathbb{T})$
we can consider the (time-independent) Floquet Hamiltonian
\begin{equation}\label{eq:Floquet}
H_{\mathrm{Floq}}=H(\theta) +\frac{1}{\tau} p_\theta=
H_{\mathrm{tot}}+ H_{\mathrm{k}} \delta_P(\theta) + \frac{1}{\tau}
p_\theta ,
\end{equation}
where
\begin{equation}\label{eq:thetap}
\theta \in \left[-1/2,1/2\right), \quad
p_\theta=-i\partial_\theta, \quad \left[\theta, p_\theta\right]=i
.
\end{equation}
We get
\begin{equation}\label{eq:motiontheta}
\dot\theta=-i\left[\theta, H_{\mathrm{Floq}}\right]= 1/\tau,
\qquad \theta(t)=t/\tau,
\end{equation}
whence $\forall A\in \mathcal{H}$,
\begin{equation}\label{eq:motionA}
\dot A(t)=-i\left[A(t), H_{\mathrm{Floq}}\right]= -i\left[A(t),
H\left(t/\tau\right)\right] ,
\end{equation}
so that every observable in $\mathcal{H}$ evolves according to the
original Hamiltonian (\ref{eq:Hamk}). The eigenvalue equation for
$p_\theta$ reads
\begin{equation}\label{eq:peigenvalues}
p_\theta |m\rangle= 2\pi m |m\rangle, \qquad
\langle\theta|m\rangle= e^{i 2\pi m\theta}, \quad m\in\mathbb{Z} .
\end{equation}
The Hamiltonian (\ref{eq:Floquet}) in $\mathcal{H}\otimes
L^2(\mathbb{T})$  represents a control by a strong continuous
coupling, analogous to that discussed in Sec.\ \ref{sec:dyndec},
if one identifies $K=1/\tau$ and $H_c=p_\theta$. Therefore, from
Eq.\ (\ref{eq:1/KExp1}) and (\ref{eq:peigenvalues}) we obtain
\begin{eqnarray}
\omega_{mn}(\tau) &=& \frac{1}{\tau} \Omega_m + \Omega_{mn}^{(1)}
+ {\rm O}\left(\tau\right) =\frac{2\pi m}{\tau} +
\Omega_{mn}^{(1)} + {\rm O}\left(\tau\right)\ ,
\label{eq:1/tauExp1}
\end{eqnarray}
and from Eq.\ (\ref{eq:gammastrongdef}) we get
\begin{eqnarray}
\gamma_{mn}^{\mathrm{k}}(\tau)= 2\pi
\kappa_n^\beta\left(\omega_{mn}(\tau)\right)= 2\pi
\kappa_n^\beta\left(\frac{2\pi m}{\tau}
+\Omega^{(1)}_{mn}+\mathrm{O}(\tau)\right)\ ,
\label{eq:gammakickA3}
\end{eqnarray}
which is Eq.~(\ref{eq:gammakick}) of the text.

\end{document}